\documentclass[10pt,%
superscriptaddress,
 amsmath,amssymb,
 twocolumn,
 aps,
prl,
]{revtex4-2}

\usepackage{graphicx}
\usepackage{dcolumn}
\usepackage{bm}
\usepackage{lineno}
\usepackage{xcolor}
\usepackage{array}

\begin{document}

\title{{\color{black}Structural origin of resonant diffraction in RuO$_2$}}

\author{Connor A. Occhialini}
\email{co2625@columbia.edu}
\affiliation{National Synchrotron Light Source II, Brookhaven National Laboratory, Upton, NY 11973, USA}
\affiliation{Department of Physics, Columbia University, New York, NY 10027, USA}

\author{Christie Nelson}
\affiliation{National Synchrotron Light Source II, Brookhaven National Laboratory, Upton, NY 11973, USA}

\author{Alessandro Bombardi}
\affiliation{Diamond Light Source, Harwell Science and Innovation Campus, Didcot, Oxfordshire OX11 0DE, United Kingdom}

\author{Shiyu Fan}
\affiliation{National Synchrotron Light Source II, Brookhaven National Laboratory, Upton, NY 11973, USA}

\author{Raul Acevedo-Esteves}
\affiliation{National Synchrotron Light Source II, Brookhaven National Laboratory, Upton, NY 11973, USA}

\author{Riccardo Comin}
\affiliation{Department of Physics, Massachusetts Institute of Technology, Cambridge, Massachusetts 02139, United States}

\author{Dmitri N. Basov}
\affiliation{Department of Physics, Columbia University, New York, NY 10027, USA}

\author{Maki Musashi}
\affiliation{Department of Applied Physics and Quantum-Phase Electronics Center (QPEC), University of Tokyo, 113-8656 Tokyo, Japan}

\author{Masashi Kawasaki}
\affiliation{Department of Applied Physics and Quantum-Phase Electronics Center (QPEC), University of Tokyo, 113-8656 Tokyo, Japan}
\affiliation{Center for Emergent Matter Science (CEMS), RIKEN, 351-0198 Saitama, Japan}

\author{Masaki Uchida}
\affiliation{Department of Physics, Institute of Science Tokyo, 152-8551 Tokyo, Japan}
\affiliation{Toyota Physical and Chemical Research Institute, 480-1192 Nagakute, Japan}

\author{Hoydoo You}
\affiliation{Materials Science Division, Argonne National Laboratory, Lemont, IL 60439, USA}

\author{John Mitchell}
\affiliation{Materials Science Division, Argonne National Laboratory, Lemont, IL 60439, USA}

\author{Valentina Bisogni}
\affiliation{National Synchrotron Light Source II, Brookhaven National Laboratory, Upton, NY 11973, USA}

\author{Claudio Mazzoli}
\affiliation{National Synchrotron Light Source II, Brookhaven National Laboratory, Upton, NY 11973, USA}

\author{Jonathan Pelliciari}
\email{pelliciari@bnl.gov}
\affiliation{National Synchrotron Light Source II, Brookhaven National Laboratory, Upton, NY 11973, USA}

\date{\today}

\begin{abstract} 
We report Ru $L_3$-edge resonant X-ray diffraction studies on single crystal and (001) epitaxial films of RuO$_2$. We investigate the distinct $\mathbf{Q} = (100)$ and $(001)$ Bragg-forbidden reflections as a function of incident energy, azimuthal angle, and temperature. The results show that the observed resonant diffraction in RuO$_2$ is fully consistent with a resonant charge anisotropy signal of structural origin permitted by the parent (non-magnetic) rutile $P4_2/mnm$ space group. These results significantly constrain the magnetic contribution to the resonant diffraction signal and indicate the unlikely existence of $\mathbf{k} = 0$ antiferromagnetic order in RuO$_2$.
\end{abstract}

\maketitle

{\it Introduction.} Collinear antiferromagnets (AFMs) exhibiting non-relativistic spin splitting (NRSS) and broken time-reversal symmetry are of interest owing to their promise for AFM spintronics  \cite{Yuan2020,Hayami2019}. Also known as altermagnets (AMs) \cite{Smejkal2022}, such materials host useful functional properties including longitudinally-polarized spin currents \cite{GonzalezHernandez2021}, an anomalous Hall effect \cite{Smejkal2020}, and chiral magnon splitting \cite{Smejkal2023, Liu2024}. The theoretical development of NRSS AFMs accelerated following reports of anomalous AFM order in metallic RuO$_2$ \cite{Berlijn2017}. RuO$_2$ has thus been a testing ground to verify AM properties from theory \cite{Smejkal2020, Smejkal2023, GonzalezHernandez2021,Hariki2024,McClarty2024} and experiments \cite{Bai2023,Fedchenko2024,Bose2022}. Nonetheless, recent studies have questioned the existence of long-range magnetic order -- in any form -- in bulk RuO$_2$ \cite{Hiraishi2024}. 

\begin{figure}
\centering

\includegraphics[width = \columnwidth]{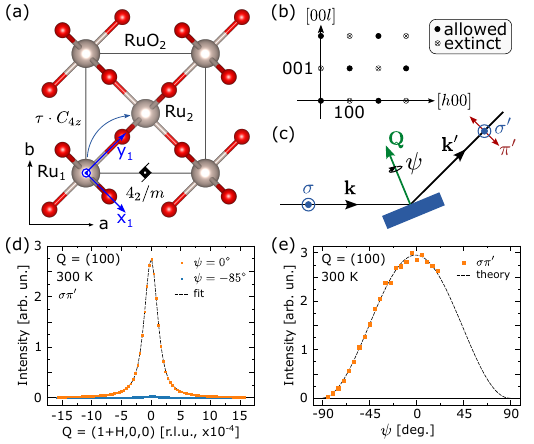}

\caption{(a) Crystal structure of RuO$_2$ along the $c$-axis. (b) Forbidden reflections for the $P4_2/mnm$ space group in the $(h0l)$ plane. (c) Scattering geometry for RXD experiments. (d) $H$ scans at $\mathbf{Q} = (100)$ in bulk RuO$_2$ for different $\psi$ in the $\sigma\pi'$ channel at $T = 300$ K. (e) $\psi$ dependence of $\mathbf{Q} = (100)$ in the same condition. Dashed lines in (d) are pseudo-Voigt fits. Dashed line in (e) is a fit to the theoretical $\psi$ dependence for both the $c$-axis AFM and ATS contributions (see text).}

\label{fig:fig1}
\end{figure}

RuO$_2$ is a metallic oxide with the rutile structure (space group $P4_2/mnm$) \cite{Mattheiss1976,Boman1970} [Fig. \ref{fig:fig1}(a,b)]. It has been of primary interest for its chemical stability, good electrical conductivity, and surface catalytic properties for oxygen evolution reaction applications \cite{Lee2012}. Further, epitaxial strain induces superconductivity in thin films \cite{Ruf2021,Uchida2020} and the characteristic non-symmorphic symmetry enforces a nodal line band structure, potentially relevant for spin Hall effect applications \cite{Jovic2018}.

Regarding the magnetic properties, early works concluded that RuO$_2$ is a Pauli paramagnet at all temperatures \cite{Ryden1970,Passenheim1969,Mattheiss1976,Glassford1994}. However, recent neutron diffraction experiments observed a structurally-forbidden reflection at $\mathbf{Q} = (100)$ which was attributed to anomalous itinerant $\mathbf{k} = 0$ AFM order with transition temperature $T_N > 300$ K \cite{Berlijn2017} [Fig. \ref{fig:fig1}(b)] ($\mathbf{Q}$ - reciprocal lattice vector, $\mathbf{k}$ - ordering wavevector). This was supported by Ru L$_2$-edge resonant X-ray diffraction (RXD), extending the evidence for AFM order to thin films by using the high sensitivity of RXD to small sample volumes \cite{Zhu2019}. However, more recent muon spin-rotation ($\mu$SR) experiments found no evidence for magnetic order \cite{Hiraishi2024} and the neutron diffraction signal was attributed to multiple scattering (non-magnetic) \cite{Kesler2024, Kiefer2024}. Multiple studies subsequently reevaluated the physical properties of RuO$_2$ using various techniques \cite{Kiefer2024,Pawula2024,Smolyanyuk2023,Wenzel2024,Yumnam2025,Paul2025,Liu2024b,Wang2025}, consistently concluding that RuO$_2$ is possibly a weak or moderately-correlated, non-magnetic metal. As the primary remaining measurement of the magnetic order parameter with sensitivity to thin films, it is thus crucial to conclusively determine whether RXD evidences magnetic order in RuO$_2$.

We report Ru L$_3$-edge ($2p \to 4d$) RXD in bulk single crystal and (001) oriented films of RuO$_2$. The previously reported RXD signal at $\mathbf{Q} = (100)$ -- attributed to magnetic order -- is instead determined to be consistent with anisotropic tensor of susceptibility (ATS) \cite{Dmitrienko1983}, or Templeton-Templeton (TT) scattering \cite{Templeton1980,Templeton1982}. ATS scattering describes a structural mechanism through which Bragg-forbidden reflections (by, e.g., glide planes or screw axes) can become allowed on resonance due to charge anisotropy of the resonant ions \cite{Dmitrienko1983}. This assignment is determined by investigation of a distinct forbidden reflection $\mathbf{Q} = (001)$ where magnetic and ATS contributions can be disentangled by distinct azimuthal ($\psi$) angle dependence, and is further supported by numerical calculations of the ATS resonance spectrum. These results significantly constrain the magnetic contribution to RXD in RuO$_2$, yielding an estimated upper bound of $|\mathbf{m}| < 0.1$ $\mu_B$ on the $\mathbf{k} = 0$ ordered magnetic moment of both bulk and (001) thin films of RuO$_2$. Combined with the conclusions of recent reports \cite{Hiraishi2024,Kesler2024,Kiefer2024,Yumnam2025}, these results indicate the likely absence of $\mathbf{k} = 0$ AFM order in RuO$_2$. 

{\it Results.} Ru L$_3$-edge RXD experiments were performed at the 4-ID (ISR) beamline at the National Synchrotron Light Source II, Brookhaven National Laboratory and at the I16 beamline at the Diamond Light Source. Bulk measurements were performed on single crystals with natural/polished (100)/(001) facets, respectively. Thin film measurements were performed on an 18-nm thick (001) RuO$_2$/TiO$_2$ epitaxial film grown by molecular beam epitaxy (MBE) \cite{Uchida2020, Occhialini2022}. The scattering geometry is depicted in Fig. \ref{fig:fig1}(c). We used incident $\sigma$ linear polarization for all measurements and resolved the scattered linear polarization $\sigma'/\pi'$ using the (002) reflection of a graphite analyzer. All data is recorded near the main Ru $L_3$ edge (incident energy $E_i \sim$ 2.838 keV) unless explicitly stated. 

We first studied $\mathbf{Q} = (100)$ [Fig. \ref{fig:fig1}(a,b)], a Bragg-forbidden reflection where Ru L$_2$-edge RXD was previously reported \cite{Zhu2019, Gregory2022}. A corresponding resonant reflection is detected at the Ru $L_3$-edge, which is present in the $\sigma\pi'$ (rotated) polarization channel even at $T = 300$ K. The $\psi$ dependence is reported in Fig. \ref{fig:fig1}(e), with $\psi = 0^\circ$ defined with the $c$-axis in the scattering plane [Fig. \ref{fig:fig1}(c)].

The $\psi$-dependence follows a nearly ideal $I(\psi) \propto\cos^2\psi$ functional form [Fig. \ref{fig:fig1}(e)]. As previously noted \cite{Zhu2019, Lovesey2022}, this is equivalently consistent with either ATS or magnetic scattering with local moments $\mathbf{m} \parallel \hat{c}$. The atomic scattering tensors in the local Ru$_1$ basis [$x_1$-$y_1$, Fig. \ref{fig:fig1}(a)] for the $D_{2h}$ charge anisotropy ($f_\mathrm{ch}$)
and isotropic magnetic scattering with $\mathbf{m} \parallel \hat{c}$ ($f_\mathrm{mag}$) contributions are \cite{Matteo2012}, respectively,
\begin{equation}
f_\mathrm{ch} \propto \begin{pmatrix} f_{xx} & \cdot & \cdot \\ \cdot & f_{yy} & \cdot \\ \cdot & \cdot & f_{zz} \end{pmatrix}, \quad 
f_\mathrm{mag} \propto i \begin{pmatrix} \cdot &   f_{xy} & \cdot \\ -f_{xy} & \cdot & \cdot \\ \cdot & \cdot & \cdot \end{pmatrix}
\label{eq:eq1}
\end{equation}

\noindent where $f_{ii}$ are unequal.

Calculations of the Ru structure factor at $\mathbf{Q} = (100)$ considering both $f_\mathrm{ch}$ and $f_\mathrm{mag}$ yields: $|F_{\sigma\pi'}|^2 \propto (|\Delta|^2 + |f_{xy}|^2 + 2 \Im(f_{xy}^* \Delta))\cos^2(\psi)$ and $|F_{\sigma\sigma'}|^2 = 0$, where $\Delta \equiv f_{xx} - f_{yy}$ \cite{suppref}. Therefore, both $c$-axis AFM and ATS scattering at $\mathbf{Q} = (100)$ appear exclusively in the cross-polarized ($\sigma\pi'$) channel and yield the same $\psi$-dependence, which agrees with the experiment [Fig. \ref{fig:fig1}(e)]. The origin of the signal cannot be uniquely determined using a $\psi$ scan with incident $\sigma$ polarization \cite{Lovesey2022}. Despite this ambiguity, we know that a non-zero ATS contribution is certain ($\Delta \neq 0$ \cite{Occhialini2021})-- as observed in isostructural IrO$_2$ \cite{Hirata2013}, TiO$_2$ \cite{Kirfel1991, Sawai2003} and MnF$_2$ \cite{Kirfel1991} -- while the existence of AFM remains uncertain \cite{Hiraishi2024}.

To isolate these contributions, we studied a distinct Bragg-forbidden reflection, $\mathbf{Q} = (001)$. This is the only other forbidden reflection accessible at the Ru L$_{2,3}$-edges. Calculation of the polarized RXD signals in the same model yields -- for {\it pure} ATS scattering: $|F_{\sigma\sigma'}|^2 \propto |\Delta|^2 \sin^2(2\psi)$ and $|F_{\sigma\pi'}|^2 \propto |\Delta|^2 \sin^2(\theta_b)\cos^2(2\psi)$; for {\it pure} magnetic scattering: $|F_{\sigma\sigma'}|^2 = 0$ and $|F_{\sigma\pi'}|^2 \propto |f_{xy}|^2$, with $\psi = 0$ defined with the $a$-axis in the scattering plane. Thus, $c$-axis AFM and ATS scattering are clearly distinguishable at $\mathbf{Q} =(001)$ \footnote{In addition, a time-odd charge-magnetic interference term $\propto \cos(2\psi)$ occurs in the $|F_{\sigma\pi'}|^2$ channel which would be observable in single AFM domain samples. We neglect this term here as it does not affect the main conclusions. Further details are provided in the SM \cite{suppref}.}.

\begin{figure}
\centering

\includegraphics[width = \columnwidth]{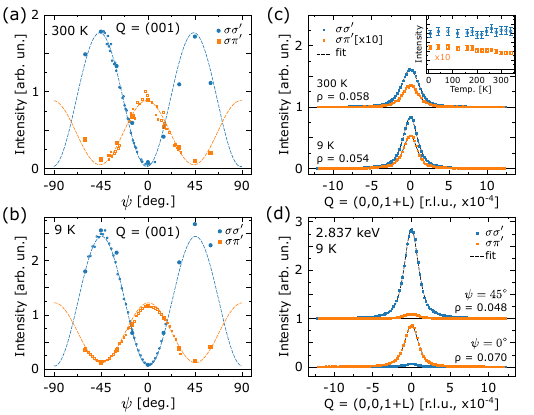}

\caption{$\psi$ dependence of $\mathbf{Q} = (001)$ in bulk RuO$_2$ in the $\sigma\sigma'$ (blue) and $\sigma\pi'$ (orange) channels at (a) $T = 300$ K and (b) $T = 9$ K. (c) $L$ scans of $\mathbf{Q} = (001)$ at $\psi = 45^\circ$ and $T = 9/300$ K. Inset: integrated $\sigma\sigma'$ and $\sigma\pi'$ intensity vs. $T$ at the same condition. (d) $L$ scans in $\sigma\sigma'$ and $\sigma\pi'$ channels at $T = 9$ K for $\psi = 0^\circ$ and $\psi = 45^\circ$. Dashed lines in (a)/(b) are fits to the theoretical $\psi$ dependence for ATS scattering considering the extinction ratio of the analyzer (see text). Black lines in (c)/(d) are pseudo-Voigt fits.}
\label{fig:fig2}
\end{figure}

Figure \ref{fig:fig2} reports our Ru L$_3$-edge RXD results at $\mathbf{Q} = (001)$. The characteristic $\psi$-dependence [Fig. \ref{fig:fig2}(a,b)] with four-fold symmetry in both $\sigma\sigma'$ and $\sigma\pi'$  allows us to immediately conclude that the RXD signal is dominated by ATS scattering. To quantify this, we simultaneously fit the full $\psi$-dependence in both polarization channels to a model including the ATS and AFM contributions, along with the extinction ratio of the polarization analyzer. At both $T = 9$ K and $T = 300$ K, the fitted $\psi$ dependence is consistent with the ATS contribution only while considering an analyzer leakage of  $\sim 5\%$.

We attempt to identify a potentially weak AFM signal by leveraging the distinct $\psi$-dependence. Maximal sensitivity to AFM occurs at $\psi = \pm 45^\circ$ where $\sigma\pi'$ ATS scattering is suppressed. Any residual $\sigma\pi'$ signal may be possibly magnetic, with sensitivity limited by the analyzer extinction ratio due to the large $\sigma\sigma'$ ATS component [Fig. \ref{fig:fig2}]. To quantify this, we define the intensity ratio between the nominally extinct and dominant polarization channels: $\rho = I_{\sigma\pi'}/I_{\sigma\sigma'}$ at $\psi = \pm 45^\circ$ and $I_{\sigma\sigma'}/I_{\sigma\pi'}$ at $\psi = 0^\circ$. Polarization resolved $L$ scans at different $\psi$ and temperatures are reported in Figs. \ref{fig:fig2}(c,d). At $T = 9$ K, we find $\rho \sim 7\%$/$4.8\%$ at $\psi = 0/45^\circ$, respectively [Fig. \ref{fig:fig2}(d)]. Thus, the  $\sigma\pi'$ channel at $\psi = 45^\circ$ displays no significant residual signal that can be attributed to magnetism above the analyzer leakage. We also measured the scattered polarization ratio at $\psi = 45^\circ$ versus temperature [Fig. \ref{fig:fig2}(c)]. No significant variation is observed in either $\rho$ or the total scattered $\sigma\sigma'$ intensity between $T = 9$ K and $300$ K. A weak decrease in the $\sigma\pi'$ component is observed in continuous $T$ scans [inset, Fig. \ref{fig:fig2}(c)], but is not significant within the systematic error introduced by the graphite analyzer homogeneity.

\begin{figure}
\centering

\includegraphics[width = \columnwidth]{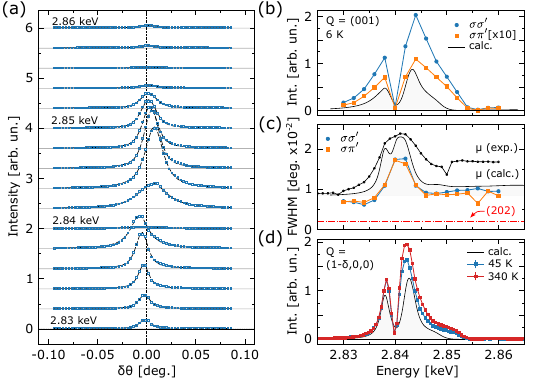}

\caption{(a) $\theta$ scans at $\mathbf{Q} = (001)$ in the $\sigma\sigma'$ channel vs. $E_i$ at $\psi = 45^\circ$ and $T = 6.5$ K, expressed as the deviation from the nominal Bragg angle ($\delta\theta$). (b) Integrated $\theta$ scans vs. $E_i$ for $\sigma\sigma'$ (blue) and $\sigma\pi'$ (orange). Black line: non-magnetic FDMNES simulation of the (001) resonance profile, accounting for self-absorption. (c) Peak width vs. $E_i$, compared to the FY XAS ($\mu$ exp., black dots) and the FDMNES calculated XAS ($\mu$ calc., black line). $\mu$ -- arbitrary units. (d) Resonance spectra at $\mathbf{Q} = (0.9975, 0, 0)$ in the $\sigma\pi'$ channel at $T = 45/340$ K (blue/red) in comparison to the non-magnetic FDMNES calculation excluding self-absorption.}

\label{fig:fig3}
\end{figure}

These results leave open the possibility for a weak AFM component which does not significantly evolve between $T = 9-300$ K. This scenario can be alternatively assessed by considering the resonance spectrum. To this end, we report rocking curves vs. $E_i$ across the Ru $L_3$-edge in Fig. \ref{fig:fig3} in the $\sigma\sigma'$ channel at $T = 9$K and $\psi = 45^\circ$. Besides a resonance of the peak intensity, effects of absorption and refraction are apparent when changing $E_i$, namely an increase of the peak width and a shift of the position from the nominal Bragg angle ($\theta_B$), respectively. This highlights that one must exercise care in determining the intrinsic resonance profile.

We thus repeated the same measurements in $\sigma\pi'$ and report the resonance spectra determined from integrated rocking curves in Fig. \ref{fig:fig3}(b). The similarity of the resonance profiles for $\sigma\sigma'$ (pure ATS) and $\sigma\pi'$ (ATS leakage plus possible AFM) indicates that the total RXD signal is determined by a single spectral function  $f(\omega)$. Indeed, $f_\mathrm{ch}(\omega)$ and $f_\mathrm{mag}(\omega)$ are expected to be distinct \cite{suppref}, with magnetic scattering preferentially resonating at the $t_{2g}$-orbital-derived peak near 2.838 keV, as observed in isoelectronic ($4d^4$) Ca$_2$RuO$_4$ \cite{Zegkinoglou2005,Porter2018} and Ca$_3$Ru$_2$O$_7$ \cite{Bohnenbuck2008}. Note that all data (e.g. Fig. \ref{fig:fig1}/\ref{fig:fig2}) are recorded near the $t_{2g}$ resonance unless explicitly stated, and therefore at the energy with the strongest sensitivity to magnetism. 

To demonstrate that the resonant spectrum is consistent with ATS scattering, we performed calculations using FDMNES  \cite{Bunau2009} (see SM \cite{suppref}). We compare the calculated (001) resonance profile to the experimental data in Fig. \ref{fig:fig3}(b), corrected for self-absorption. The agreement to the RXD spectrum and the fluorescence yield (FY) absorption curve ($\mu$) at T = 300 K [Fig. \ref{fig:fig3}(c)] is satisfactory. The slight deviations in the $t_{2g}$-$e_g$ orbital splitting is likely due to an intermediate state multiplet effect \cite{Hu2000, Occhialini2021}. However, the reasonable agreement using single particle treatment underscores the weakly correlated nature of RuO$_2$ as expected for a $4d$ metal.

The resonance profile for pure ATS scattering is expected to be identical for the $(001)$ and $(100)$ reflections, determined by the same spectral function $f_\mathrm{ch}(\omega)$. We measured T-dependent RXD spectra near (100) at a detuned $\mathbf{Q} = (0.9975, 0, 0)$ in $\sigma\pi'$ polarization [Fig. \ref{fig:fig3}(e)]. The detuning in momentum-space approximately compensates self-absorption through the inverse effect of the peak broadening, allowing a more direct measurement of the intrinsic resonance spectrum. The $\mathbf{Q} = (100)$ RXD spectrum is nearly T-independent between $45-340$ K and compares well with the same result from FDMNES used for $\mathbf{Q} = (001)$ above, excluding self-absorption. 

Thus, the (100) and (001) resonance profiles are well-described by the same ATS spectral function, calculated using the non-magnetic $P4_2/mnm$ space group. Meanwhile, spin-polarized calculations considering a finite Ru magnetic moment and a $\mathbf{k} = 0$ AFM order confirm that magnetic scattering preferentially resonates at the $t_{2g}$ orbital peak \cite{suppref}, at odds with the present experiments. Utilizing these distinct ATS and magnetic resonance spectra and their relative intensity as predicted by FDMNES, we quantify an approximate upper bound of the $\mathbf{k} = 0$ ordered AFM moment to $|\mathbf{m}| < 0.1$ $\mu_B$ \cite{suppref}.

Separately, we note that the peak width versus $E_i$ closely follows the absorption spectrum [Fig. \ref{fig:fig3}(c)]. This suggests that the peak width is determined by the energy-dependent absorption length, which can be quantitatively described by our FDMNES calculations \cite{suppref}. 
Notably, the FWHM of the $\sigma\pi'$ and $\sigma\sigma'$ rocking curves are nearly identical, while both are broader than the $(202)$ reflection measured with the $3^{\text{rd}}$ harmonic [8.514 keV, dashed line Fig. \ref{fig:fig3}(c)]. This precludes an attribution of the increased peak width on resonance to a finite magnetic correlation length. A corresponding analysis at $\mathbf{Q} = (100)$ reveals quantitatively similar behavior \cite{suppref}.

\begin{figure}[tb]
\centering

\includegraphics[width = \columnwidth]{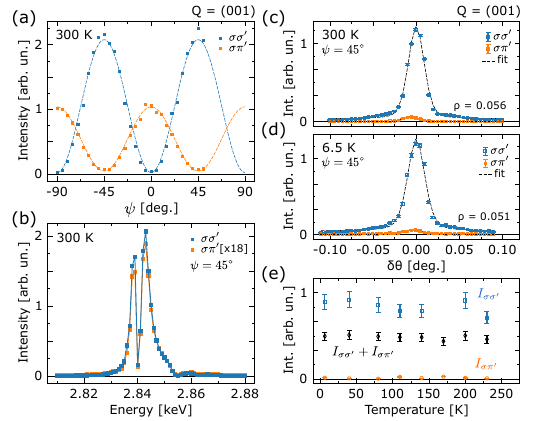}

\caption{(a) $\psi$ dependence of the (001) reflection in an epitaxial (001) RuO$_2$/TiO$_2$ film at $T = 300$ K. (b) Energy scans at $(001)$ in the $\sigma\sigma'$ (blue) and $\sigma\pi'$ (orange) polarizations at $\psi = 45^\circ$. $\theta$ scans of the (001) reflection in both polarizations at (c) $T = 300$ K and (d) $T = 6.5$ K at $\psi = 45^\circ$. (e) $T$ dependence of each polarization channel ($I_{\sigma\sigma'} + I_{\sigma\pi'}$ -- without analyzer).}

\label{fig:fig4}
\end{figure}

Having established that the RXD signal in bulk RuO$_2$ is consistent with ATS scattering, we apply the same technique to high-quality epitaxial (001) RuO$_2$/TiO$_2$ thin films grown by MBE \cite{Uchida2020}. We performed the same set of measurements at $\mathbf{Q} = (001)$ as in the bulk, as reported in Figure \ref{fig:fig4}. The thin film results support the same conclusions: (i), the $\psi$-dependence is consistent with pure ATS scattering at $T = 300$ K considering $5\%$ analyzer leakage [Fig. \ref{fig:fig4}(a)]; (ii), the resonance spectra at $\psi = 45^\circ$ are nearly identical in $\sigma\sigma'$ and $\sigma\pi'$, indicating a single component spectral function [Fig. \ref{fig:fig4}(b)], which agrees closely with the bulk result and expectations from non-magnetic FDMNES calculations; (iii), the $T$ dependence at $\psi = 45^\circ$, which maximizes sensitivity to any magnetic component, shows that the polarization ratio $\rho \sim 0.05$ is nearly $T$-independent between $6.5-300$ K [Fig. \ref{fig:fig4}(c,d)]; and (iv), the (001) reflection in all polarization channels is at most weakly dependent on $T$ [Fig. \ref{fig:fig4}(e)]. Note that details of the refraction and absorption which complicate the bulk measurements [Fig. \ref{fig:fig3}] are not significant in the thin film since the absorption length ($\geq 400$ nm) exceeds the film thickness ($\sim 18$ nm). This enables a clean and direct characterization of the (001) RXD signal, mutually supporting the conclusion of the absence of $\mathbf{k} = 0$ AFM order as described for bulk RuO$_2$. 

{\it Discussion.} We conclude that Ru $L$-edge RXD in RuO$_2$ is dominated by ATS scattering of structural origin, consistent with the non-magnetic $P4_2/mnm$ space group. Any magnetic contribution, if present at all, is challenging to extract from this large structural contribution. Based on the experimental sensitivity of our RXD results, we estimate a bound on the $\mathbf{k} = 0$ ordered AFM moment to $|\mathbf{m}| < 0.1 \mu_B$. In the context of recent $\mu$SR \cite{Hiraishi2024} ($|\mathbf{m}| < 4.8 \times 10^{-4}$ $\mu_B$), neutron scattering \cite{Kesler2024,Kiefer2024} ($|\mathbf{m}| < 0.01$ $\mu_B$), and M\"{o}ssbauer \cite{Yumnam2025} ($|\mathbf{m}| < 0.03$ $\mu_B$) studies, the most likely conclusion of the presented results is that bulk RuO$_2$ is non-magnetic. Our results demonstrate the same conclusion for high-quality (001) epitaxial thin films.

The difficulties of probing AFM order with RXD highlighted here for RuO$_2$ are potentially common among other AM candidates due to close similarities in the defining symmetries of AM \cite{Smejkal2022, Yuan2023} with those permitting ATS scattering \cite{Templeton1980, Templeton1982, Dmitrienko1983}. By the definition proposed by \v{S}mejkal et al. \cite{Smejkal2022}, AMs describe {\it collinear} AFMs where the two AFM sublattices are symmetry-related through a combined non-trivial rotation and translation (e.g., the $4_2$ screw axis in RuO$_2$), as opposed to a simple translation (lattice centering) or inversion \cite{Smejkal2020, Smejkal2022, Yuan2020, Yuan2023}. Thus, several lattice structures supporting AM have an internal structural sublattice symmetry [e.g. Ru$_1$/Ru$_2$, Fig. \ref{fig:fig1}(a)] present within a primitive Bravais lattice, leading to Bragg-forbidden reflections. These structural sublattices are then directly mapped to the AFM sublattices through a $\mathbf{k} = 0$ AFM order. In this situation, magnetic diffraction is expected at precisely the integral $\mathbf{Q}$ positions where non-resonant structural contributions from the two sublattices cancel yielding a forbidden reflection, which may be relaxed by ATS scattering due to the same inter-sublattice symmetry operation inducing AM \cite{Dmitrienko1983}.

These properties may complicate the study of AM with RXD in certain lattice symmetries, requiring care to disentangle overlapping ATS and AFM contributions. For the rutile structure, $\mathbf{Q} = (100)$ is a special case where the RXD signals from ATS and $c$-axis AFM order overlap exactly, while they can be effectively disentangled at the distinct $\mathbf{Q} = (001)$, as we have shown. The specific protocol for other systems will inevitably depend on details of the crystallographic and magnetic structures. Interestingly, overlapping ATS and magnetic RXD contributions allow a chiral signature in RXD resulting from charge-magnetic interference as originally proposed by Lovesey et al. \cite{Lovesey2022}, effectively yielding a magnetic analog to Templeton scattering. These charge-magnetic interference terms are time-odd and thus could be a promising application of RXD for, e.g., the detection of AFM domains in AMs.

To circumvent the difficulty of probing $\mathbf{k} = 0$ magnetic order in rutile AFMs with RXD, previous studies have instead used non-resonant magnetic scattering \cite{Bergevin1972,Strempfer1996}. However, high symmetry reciprocal lattice points are also particularly sensitive to multiple scattering \cite{Strempfer1996}, which is notably the same effect which led to the non-magnetic artifact in neutron investigations \cite{Kesler2024}. Therefore, space group symmetry must be carefully considered when investigating AM candidates in diffraction experiments.

{\it Conclusion.} We demonstrated that RXD at the Ru $L$-edges in RuO$_2$ is dominated by structural effects through ATS scattering. This reconciles the apparent inconsistency between RXD and other techniques which demonstrate that RuO$_2$ is non-magnetic. \\

{\it Acknowledgments.} We acknowledge discussions with Y. Joly, B. Gregory, K. Shen, I. Mazin, and M. Braden. This work was supported as part of Programmable Quantum Materials, an Energy Frontier Research Center funded by the US Department of Energy (DOE), Office of Science, Basic Energy Sciences (BES), under award DE-SC0019443 (C.A.O., D.N.B.). This research used beamline 4-ID (ISR) of the National Synchrotron Light Source II, a U.S. Department of Energy (DOE) Office of Science User Facility operated for the DOE Office of Science by Brookhaven National Laboratory under Contract No. DE-SC0012704. This work was carried out with the support of Diamond Light Source, instrument I16 (proposal MM40048). We thank A. Abdeldaim and R. Scatena for technical assistance during the experiments at Diamond Light Source beamline I16. This work was supported by JST FOREST Program Grant Number JPMJFR202N, by JSPS KAKENHI Grant Numbers JP24H01654 and JP24H01614 from MEXT, Japan, by Toyota Riken Rising Fellow Program funded by Toyota Physical and Chemical Research Institute, Japan, and by STAR Award funded by the Tokyo Tech Fund, Japan.

\appendix


\begin{thebibliography}{57}%
\makeatletter
\providecommand \@ifxundefined [1]{%
 \@ifx{#1\undefined}
}%
\providecommand \@ifnum [1]{%
 \ifnum #1\expandafter \@firstoftwo
 \else \expandafter \@secondoftwo
 \fi
}%
\providecommand \@ifx [1]{%
 \ifx #1\expandafter \@firstoftwo
 \else \expandafter \@secondoftwo
 \fi
}%
\providecommand \natexlab [1]{#1}%
\providecommand \enquote  [1]{``#1''}%
\providecommand \bibnamefont  [1]{#1}%
\providecommand \bibfnamefont [1]{#1}%
\providecommand \citenamefont [1]{#1}%
\providecommand \href@noop [0]{\@secondoftwo}%
\providecommand \href [0]{\begingroup \@sanitize@url \@href}%
\providecommand \@href[1]{\@@startlink{#1}\@@href}%
\providecommand \@@href[1]{\endgroup#1\@@endlink}%
\providecommand \@sanitize@url [0]{\catcode `\\12\catcode `\$12\catcode `\&12\catcode `\#12\catcode `\^12\catcode `\_12\catcode `\%12\relax}%
\providecommand \@@startlink[1]{}%
\providecommand \@@endlink[0]{}%
\providecommand \url  [0]{\begingroup\@sanitize@url \@url }%
\providecommand \@url [1]{\endgroup\@href {#1}{\urlprefix }}%
\providecommand \urlprefix  [0]{URL }%
\providecommand \Eprint [0]{\href }%
\providecommand \doibase [0]{https://doi.org/}%
\providecommand \selectlanguage [0]{\@gobble}%
\providecommand \bibinfo  [0]{\@secondoftwo}%
\providecommand \bibfield  [0]{\@secondoftwo}%
\providecommand \translation [1]{[#1]}%
\providecommand \BibitemOpen [0]{}%
\providecommand \bibitemStop [0]{}%
\providecommand \bibitemNoStop [0]{.\EOS\space}%
\providecommand \EOS [0]{\spacefactor3000\relax}%
\providecommand \BibitemShut  [1]{\csname bibitem#1\endcsname}%
\let\auto@bib@innerbib\@empty
\bibitem [{\citenamefont {Yuan}\ \emph {et~al.}(2020)\citenamefont {Yuan}, \citenamefont {Wang}, \citenamefont {Luo}, \citenamefont {Rashba},\ and\ \citenamefont {Zunger}}]{Yuan2020}%
  \BibitemOpen
  \bibfield  {author} {\bibinfo {author} {\bibfnamefont {L.~D.}\ \bibnamefont {Yuan}}, \bibinfo {author} {\bibfnamefont {Z.}~\bibnamefont {Wang}}, \bibinfo {author} {\bibfnamefont {J.~W.}\ \bibnamefont {Luo}}, \bibinfo {author} {\bibfnamefont {E.~I.}\ \bibnamefont {Rashba}},\ and\ \bibinfo {author} {\bibfnamefont {A.}~\bibnamefont {Zunger}},\ }\bibfield  {title} {\bibinfo {title} {{Giant momentum-dependent spin splitting in centrosymmetric low-Z antiferromagnets}},\ }\href {https://doi.org/10.1103/PhysRevB.102.014422} {\bibfield  {journal} {\bibinfo  {journal} {Physical Review B}\ }\textbf {\bibinfo {volume} {102}},\ \bibinfo {pages} {014422} (\bibinfo {year} {2020})}\BibitemShut {NoStop}%
\bibitem [{\citenamefont {Hayami}\ \emph {et~al.}(2019)\citenamefont {Hayami}, \citenamefont {Yanagi},\ and\ \citenamefont {Kusunose}}]{Hayami2019}%
  \BibitemOpen
  \bibfield  {author} {\bibinfo {author} {\bibfnamefont {S.}~\bibnamefont {Hayami}}, \bibinfo {author} {\bibfnamefont {Y.}~\bibnamefont {Yanagi}},\ and\ \bibinfo {author} {\bibfnamefont {H.}~\bibnamefont {Kusunose}},\ }\bibfield  {title} {\bibinfo {title} {Momentum-dependent spin splitting by collinear antiferromagnetic ordering},\ }\href {https://doi.org/10.7566/JPSJ.88.123702} {\bibfield  {journal} {\bibinfo  {journal} {Journal of the Physical Society of Japan}\ }\textbf {\bibinfo {volume} {88}},\ \bibinfo {pages} {123702} (\bibinfo {year} {2019})}\BibitemShut {NoStop}%
\bibitem [{\citenamefont {Šmejkal}\ \emph {et~al.}(2022)\citenamefont {Šmejkal}, \citenamefont {Sinova},\ and\ \citenamefont {Jungwirth}}]{Smejkal2022}%
  \BibitemOpen
  \bibfield  {author} {\bibinfo {author} {\bibfnamefont {L.}~\bibnamefont {Šmejkal}}, \bibinfo {author} {\bibfnamefont {J.}~\bibnamefont {Sinova}},\ and\ \bibinfo {author} {\bibfnamefont {T.}~\bibnamefont {Jungwirth}},\ }\bibfield  {title} {\bibinfo {title} {{Beyond Conventional Ferromagnetism and Antiferromagnetism: A Phase with Nonrelativistic Spin and Crystal Rotation Symmetry}},\ }\href {https://doi.org/10.1103/PhysRevX.12.031042} {\bibfield  {journal} {\bibinfo  {journal} {Physical Review X}\ }\textbf {\bibinfo {volume} {12}},\ \bibinfo {pages} {031042} (\bibinfo {year} {2022})}\BibitemShut {NoStop}%
\bibitem [{\citenamefont {González-Hernández}\ \emph {et~al.}(2021)\citenamefont {González-Hernández}, \citenamefont {Šmejkal}, \citenamefont {Výborný}, \citenamefont {Yahagi}, \citenamefont {Sinova}, \citenamefont {Jungwirth},\ and\ \citenamefont {Železný}}]{GonzalezHernandez2021}%
  \BibitemOpen
  \bibfield  {author} {\bibinfo {author} {\bibfnamefont {R.}~\bibnamefont {González-Hernández}}, \bibinfo {author} {\bibfnamefont {L.}~\bibnamefont {Šmejkal}}, \bibinfo {author} {\bibfnamefont {K.}~\bibnamefont {Výborný}}, \bibinfo {author} {\bibfnamefont {Y.}~\bibnamefont {Yahagi}}, \bibinfo {author} {\bibfnamefont {J.}~\bibnamefont {Sinova}}, \bibinfo {author} {\bibfnamefont {T.}~\bibnamefont {Jungwirth}},\ and\ \bibinfo {author} {\bibfnamefont {J.}~\bibnamefont {Železný}},\ }\bibfield  {title} {\bibinfo {title} {Efficient electrical spin splitter based on nonrelativistic collinear antiferromagnetism},\ }\href {https://doi.org/10.1103/PhysRevLett.126.127701} {\bibfield  {journal} {\bibinfo  {journal} {Physical Review Letters}\ }\textbf {\bibinfo {volume} {126}},\ \bibinfo {pages} {127701} (\bibinfo {year} {2021})}\BibitemShut {NoStop}%
\bibitem [{\citenamefont {Šmejkal}\ \emph {et~al.}(2020)\citenamefont {Šmejkal}, \citenamefont {González-Hernández}, \citenamefont {Jungwirth},\ and\ \citenamefont {Sinova}}]{Smejkal2020}%
  \BibitemOpen
  \bibfield  {author} {\bibinfo {author} {\bibfnamefont {L.}~\bibnamefont {Šmejkal}}, \bibinfo {author} {\bibfnamefont {R.}~\bibnamefont {González-Hernández}}, \bibinfo {author} {\bibfnamefont {T.}~\bibnamefont {Jungwirth}},\ and\ \bibinfo {author} {\bibfnamefont {J.}~\bibnamefont {Sinova}},\ }\bibfield  {title} {\bibinfo {title} {Crystal time-reversal symmetry breaking and spontaneous hall effect in collinear antiferromagnets},\ }\href@noop {} {\bibfield  {journal} {\bibinfo  {journal} {Sci. Adv}\ }\textbf {\bibinfo {volume} {6}},\ \bibinfo {pages} {eaaz8809} (\bibinfo {year} {2020})}\BibitemShut {NoStop}%
\bibitem [{\citenamefont {Šmejkal}\ \emph {et~al.}(2023)\citenamefont {Šmejkal}, \citenamefont {Marmodoro}, \citenamefont {Ahn}, \citenamefont {González-Hernández}, \citenamefont {Turek}, \citenamefont {Mankovsky}, \citenamefont {Ebert}, \citenamefont {D'Souza}, \citenamefont {Šipr}, \citenamefont {Sinova},\ and\ \citenamefont {Jungwirth}}]{Smejkal2023}%
  \BibitemOpen
  \bibfield  {author} {\bibinfo {author} {\bibfnamefont {L.}~\bibnamefont {Šmejkal}}, \bibinfo {author} {\bibfnamefont {A.}~\bibnamefont {Marmodoro}}, \bibinfo {author} {\bibfnamefont {K.~H.}\ \bibnamefont {Ahn}}, \bibinfo {author} {\bibfnamefont {R.}~\bibnamefont {González-Hernández}}, \bibinfo {author} {\bibfnamefont {I.}~\bibnamefont {Turek}}, \bibinfo {author} {\bibfnamefont {S.}~\bibnamefont {Mankovsky}}, \bibinfo {author} {\bibfnamefont {H.}~\bibnamefont {Ebert}}, \bibinfo {author} {\bibfnamefont {S.~W.}\ \bibnamefont {D'Souza}}, \bibinfo {author} {\bibfnamefont {O.}~\bibnamefont {Šipr}}, \bibinfo {author} {\bibfnamefont {J.}~\bibnamefont {Sinova}},\ and\ \bibinfo {author} {\bibfnamefont {T.}~\bibnamefont {Jungwirth}},\ }\bibfield  {title} {\bibinfo {title} {{Chiral Magnons in Altermagnetic RuO$_2$}},\ }\href {https://doi.org/10.1103/PhysRevLett.131.256703} {\bibfield  {journal} {\bibinfo  {journal} {Physical Review Letters}\ }\textbf {\bibinfo {volume} {131}},\ \bibinfo {pages} {256703} (\bibinfo
  {year} {2023})}\BibitemShut {NoStop}%
\bibitem [{\citenamefont {Liu}\ \emph {et~al.}(2024{\natexlab{a}})\citenamefont {Liu}, \citenamefont {Ozeki}, \citenamefont {Asai}, \citenamefont {Itoh},\ and\ \citenamefont {Masuda}}]{Liu2024}%
  \BibitemOpen
  \bibfield  {author} {\bibinfo {author} {\bibfnamefont {Z.}~\bibnamefont {Liu}}, \bibinfo {author} {\bibfnamefont {M.}~\bibnamefont {Ozeki}}, \bibinfo {author} {\bibfnamefont {S.}~\bibnamefont {Asai}}, \bibinfo {author} {\bibfnamefont {S.}~\bibnamefont {Itoh}},\ and\ \bibinfo {author} {\bibfnamefont {T.}~\bibnamefont {Masuda}},\ }\bibfield  {title} {\bibinfo {title} {{Chiral-Split Magnon in Altermagnetic MnTe}},\ }\href {https://doi.org/10.1103/PhysRevLett.133.156702} {\bibfield  {journal} {\bibinfo  {journal} {Physical Review Letters}\ }\textbf {\bibinfo {volume} {133}},\ \bibinfo {pages} {156702} (\bibinfo {year} {2024}{\natexlab{a}})}\BibitemShut {NoStop}%
\bibitem [{\citenamefont {Berlijn}\ \emph {et~al.}(2017)\citenamefont {Berlijn}, \citenamefont {Snijders}, \citenamefont {Delaire}, \citenamefont {Zhou}, \citenamefont {Maier}, \citenamefont {Cao}, \citenamefont {Chi}, \citenamefont {Matsuda}, \citenamefont {Wang}, \citenamefont {Koehler}, \citenamefont {Kent},\ and\ \citenamefont {Weitering}}]{Berlijn2017}%
  \BibitemOpen
  \bibfield  {author} {\bibinfo {author} {\bibfnamefont {T.}~\bibnamefont {Berlijn}}, \bibinfo {author} {\bibfnamefont {P.~C.}\ \bibnamefont {Snijders}}, \bibinfo {author} {\bibfnamefont {O.}~\bibnamefont {Delaire}}, \bibinfo {author} {\bibfnamefont {H.~D.}\ \bibnamefont {Zhou}}, \bibinfo {author} {\bibfnamefont {T.~A.}\ \bibnamefont {Maier}}, \bibinfo {author} {\bibfnamefont {H.~B.}\ \bibnamefont {Cao}}, \bibinfo {author} {\bibfnamefont {S.~X.}\ \bibnamefont {Chi}}, \bibinfo {author} {\bibfnamefont {M.}~\bibnamefont {Matsuda}}, \bibinfo {author} {\bibfnamefont {Y.}~\bibnamefont {Wang}}, \bibinfo {author} {\bibfnamefont {M.~R.}\ \bibnamefont {Koehler}}, \bibinfo {author} {\bibfnamefont {P.~R.}\ \bibnamefont {Kent}},\ and\ \bibinfo {author} {\bibfnamefont {H.~H.}\ \bibnamefont {Weitering}},\ }\bibfield  {title} {\bibinfo {title} {{Itinerant Antiferromagnetism in RuO$_2$}},\ }\href {https://doi.org/10.1103/PhysRevLett.118.077201} {\bibfield  {journal} {\bibinfo  {journal} {Physical Review Letters}\ }\textbf
  {\bibinfo {volume} {118}},\ \bibinfo {pages} {077201} (\bibinfo {year} {2017})}\BibitemShut {NoStop}%
\bibitem [{\citenamefont {Hariki}\ \emph {et~al.}(2024)\citenamefont {Hariki}, \citenamefont {Takahashi},\ and\ \citenamefont {Kuneš}}]{Hariki2024}%
  \BibitemOpen
  \bibfield  {author} {\bibinfo {author} {\bibfnamefont {A.}~\bibnamefont {Hariki}}, \bibinfo {author} {\bibfnamefont {Y.}~\bibnamefont {Takahashi}},\ and\ \bibinfo {author} {\bibfnamefont {J.}~\bibnamefont {Kuneš}},\ }\bibfield  {title} {\bibinfo {title} {{X-ray magnetic circular dichroism in RuO$_2$}},\ }\href {https://doi.org/10.1103/PhysRevB.109.094413} {\bibfield  {journal} {\bibinfo  {journal} {Physical Review B}\ }\textbf {\bibinfo {volume} {109}},\ \bibinfo {pages} {094413} (\bibinfo {year} {2024})}\BibitemShut {NoStop}%
\bibitem [{\citenamefont {McClarty}\ and\ \citenamefont {Rau}(2024)}]{McClarty2024}%
  \BibitemOpen
  \bibfield  {author} {\bibinfo {author} {\bibfnamefont {P.~A.}\ \bibnamefont {McClarty}}\ and\ \bibinfo {author} {\bibfnamefont {J.~G.}\ \bibnamefont {Rau}},\ }\bibfield  {title} {\bibinfo {title} {{Landau Theory of Altermagnetism}},\ }\href {https://doi.org/10.1103/PhysRevLett.132.176702} {\bibfield  {journal} {\bibinfo  {journal} {Physical Review Letters}\ }\textbf {\bibinfo {volume} {132}},\ \bibinfo {pages} {176702} (\bibinfo {year} {2024})}\BibitemShut {NoStop}%
\bibitem [{\citenamefont {Bai}\ \emph {et~al.}(2023)\citenamefont {Bai}, \citenamefont {Zhang}, \citenamefont {Zhou}, \citenamefont {Chen}, \citenamefont {Wan}, \citenamefont {Han}, \citenamefont {Zhu}, \citenamefont {Liang}, \citenamefont {Su}, \citenamefont {Han}, \citenamefont {Pan},\ and\ \citenamefont {Song}}]{Bai2023}%
  \BibitemOpen
  \bibfield  {author} {\bibinfo {author} {\bibfnamefont {H.}~\bibnamefont {Bai}}, \bibinfo {author} {\bibfnamefont {Y.~C.}\ \bibnamefont {Zhang}}, \bibinfo {author} {\bibfnamefont {Y.~J.}\ \bibnamefont {Zhou}}, \bibinfo {author} {\bibfnamefont {P.}~\bibnamefont {Chen}}, \bibinfo {author} {\bibfnamefont {C.~H.}\ \bibnamefont {Wan}}, \bibinfo {author} {\bibfnamefont {L.}~\bibnamefont {Han}}, \bibinfo {author} {\bibfnamefont {W.~X.}\ \bibnamefont {Zhu}}, \bibinfo {author} {\bibfnamefont {S.~X.}\ \bibnamefont {Liang}}, \bibinfo {author} {\bibfnamefont {Y.~C.}\ \bibnamefont {Su}}, \bibinfo {author} {\bibfnamefont {X.~F.}\ \bibnamefont {Han}}, \bibinfo {author} {\bibfnamefont {F.}~\bibnamefont {Pan}},\ and\ \bibinfo {author} {\bibfnamefont {C.}~\bibnamefont {Song}},\ }\bibfield  {title} {\bibinfo {title} {{Efficient Spin-to-Charge Conversion via Altermagnetic Spin Splitting Effect in Antiferromagnet RuO$_2$}},\ }\href {https://doi.org/10.1103/PhysRevLett.130.216701} {\bibfield  {journal} {\bibinfo  {journal}
  {Physical Review Letters}\ }\textbf {\bibinfo {volume} {130}},\ \bibinfo {pages} {216701} (\bibinfo {year} {2023})}\BibitemShut {NoStop}%
\bibitem [{\citenamefont {Fedchenko}\ \emph {et~al.}(2024)\citenamefont {Fedchenko}, \citenamefont {Minár}, \citenamefont {Akashdeep}, \citenamefont {D'souza}, \citenamefont {Vasilyev}, \citenamefont {Tkach}, \citenamefont {Odenbreit}, \citenamefont {Nguyen}, \citenamefont {Kutnyakhov}, \citenamefont {Wind}, \citenamefont {Wenthaus}, \citenamefont {Scholz}, \citenamefont {Rossnagel}, \citenamefont {Hoesch}, \citenamefont {Aeschlimann}, \citenamefont {Stadtmüller}, \citenamefont {Kläui}, \citenamefont {Schönhense}, \citenamefont {Jungwirth}, \citenamefont {Hellenes}, \citenamefont {Jakob}, \citenamefont {Šmejkal}, \citenamefont {Sinova},\ and\ \citenamefont {Elmers}}]{Fedchenko2024}%
  \BibitemOpen
  \bibfield  {author} {\bibinfo {author} {\bibfnamefont {O.}~\bibnamefont {Fedchenko}}, \bibinfo {author} {\bibfnamefont {J.}~\bibnamefont {Minár}}, \bibinfo {author} {\bibfnamefont {A.}~\bibnamefont {Akashdeep}}, \bibinfo {author} {\bibfnamefont {S.~W.}\ \bibnamefont {D'souza}}, \bibinfo {author} {\bibfnamefont {D.}~\bibnamefont {Vasilyev}}, \bibinfo {author} {\bibfnamefont {O.}~\bibnamefont {Tkach}}, \bibinfo {author} {\bibfnamefont {L.}~\bibnamefont {Odenbreit}}, \bibinfo {author} {\bibfnamefont {Q.}~\bibnamefont {Nguyen}}, \bibinfo {author} {\bibfnamefont {D.}~\bibnamefont {Kutnyakhov}}, \bibinfo {author} {\bibfnamefont {N.}~\bibnamefont {Wind}}, \bibinfo {author} {\bibfnamefont {L.}~\bibnamefont {Wenthaus}}, \bibinfo {author} {\bibfnamefont {M.}~\bibnamefont {Scholz}}, \bibinfo {author} {\bibfnamefont {K.}~\bibnamefont {Rossnagel}}, \bibinfo {author} {\bibfnamefont {M.}~\bibnamefont {Hoesch}}, \bibinfo {author} {\bibfnamefont {M.}~\bibnamefont {Aeschlimann}}, \bibinfo {author} {\bibfnamefont
  {B.}~\bibnamefont {Stadtmüller}}, \bibinfo {author} {\bibfnamefont {M.}~\bibnamefont {Kläui}}, \bibinfo {author} {\bibfnamefont {G.}~\bibnamefont {Schönhense}}, \bibinfo {author} {\bibfnamefont {T.}~\bibnamefont {Jungwirth}}, \bibinfo {author} {\bibfnamefont {A.~B.}\ \bibnamefont {Hellenes}}, \bibinfo {author} {\bibfnamefont {G.}~\bibnamefont {Jakob}}, \bibinfo {author} {\bibfnamefont {L.}~\bibnamefont {Šmejkal}}, \bibinfo {author} {\bibfnamefont {J.}~\bibnamefont {Sinova}},\ and\ \bibinfo {author} {\bibfnamefont {H.-J.}\ \bibnamefont {Elmers}},\ }\bibfield  {title} {\bibinfo {title} {{Observation of time-reversal symmetry breaking in the band structure of altermagnetic RuO$_2$}},\ }\href@noop {} {\bibfield  {journal} {\bibinfo  {journal} {Sci. Adv}\ }\textbf {\bibinfo {volume} {10}},\ \bibinfo {pages} {eadj4883} (\bibinfo {year} {2024})}\BibitemShut {NoStop}%
\bibitem [{\citenamefont {Bose}\ \emph {et~al.}(2022)\citenamefont {Bose}, \citenamefont {Schreiber}, \citenamefont {Jain}, \citenamefont {Shao}, \citenamefont {Nair}, \citenamefont {Sun}, \citenamefont {Zhang}, \citenamefont {Muller}, \citenamefont {Tsymbal}, \citenamefont {Schlom},\ and\ \citenamefont {Ralph}}]{Bose2022}%
  \BibitemOpen
  \bibfield  {author} {\bibinfo {author} {\bibfnamefont {A.}~\bibnamefont {Bose}}, \bibinfo {author} {\bibfnamefont {N.~J.}\ \bibnamefont {Schreiber}}, \bibinfo {author} {\bibfnamefont {R.}~\bibnamefont {Jain}}, \bibinfo {author} {\bibfnamefont {D.~F.}\ \bibnamefont {Shao}}, \bibinfo {author} {\bibfnamefont {H.~P.}\ \bibnamefont {Nair}}, \bibinfo {author} {\bibfnamefont {J.}~\bibnamefont {Sun}}, \bibinfo {author} {\bibfnamefont {X.~S.}\ \bibnamefont {Zhang}}, \bibinfo {author} {\bibfnamefont {D.~A.}\ \bibnamefont {Muller}}, \bibinfo {author} {\bibfnamefont {E.~Y.}\ \bibnamefont {Tsymbal}}, \bibinfo {author} {\bibfnamefont {D.~G.}\ \bibnamefont {Schlom}},\ and\ \bibinfo {author} {\bibfnamefont {D.~C.}\ \bibnamefont {Ralph}},\ }\bibfield  {title} {\bibinfo {title} {{Tilted spin current generated by the collinear antiferromagnet ruthenium dioxide}},\ }\href {https://doi.org/10.1038/s41928-022-00744-8} {\bibfield  {journal} {\bibinfo  {journal} {Nature Electronics}\ }\textbf {\bibinfo {volume} {5}},\ \bibinfo
  {pages} {267} (\bibinfo {year} {2022})}\BibitemShut {NoStop}%
\bibitem [{\citenamefont {Hiraishi}\ \emph {et~al.}(2024)\citenamefont {Hiraishi}, \citenamefont {Okabe}, \citenamefont {Koda}, \citenamefont {Kadono}, \citenamefont {Muroi}, \citenamefont {Hirai},\ and\ \citenamefont {Hiroi}}]{Hiraishi2024}%
  \BibitemOpen
  \bibfield  {author} {\bibinfo {author} {\bibfnamefont {M.}~\bibnamefont {Hiraishi}}, \bibinfo {author} {\bibfnamefont {H.}~\bibnamefont {Okabe}}, \bibinfo {author} {\bibfnamefont {A.}~\bibnamefont {Koda}}, \bibinfo {author} {\bibfnamefont {R.}~\bibnamefont {Kadono}}, \bibinfo {author} {\bibfnamefont {T.}~\bibnamefont {Muroi}}, \bibinfo {author} {\bibfnamefont {D.}~\bibnamefont {Hirai}},\ and\ \bibinfo {author} {\bibfnamefont {Z.}~\bibnamefont {Hiroi}},\ }\bibfield  {title} {\bibinfo {title} {{Nonmagnetic Ground State in RuO$_2$ Revealed by Muon Spin Rotation}},\ }\href {https://doi.org/10.1103/PhysRevLett.132.166702} {\bibfield  {journal} {\bibinfo  {journal} {Physical Review Letters}\ }\textbf {\bibinfo {volume} {132}},\ \bibinfo {pages} {166702} (\bibinfo {year} {2024})}\BibitemShut {NoStop}%
\bibitem [{\citenamefont {Mattheiss}(1976)}]{Mattheiss1976}%
  \BibitemOpen
  \bibfield  {author} {\bibinfo {author} {\bibfnamefont {L.~F.}\ \bibnamefont {Mattheiss}},\ }\bibfield  {title} {\bibinfo {title} {{Electronic structure of RuO$_2$, OsO$_2$, and IrO$_2$}},\ }\href@noop {} {\bibfield  {journal} {\bibinfo  {journal} {Physical Review B}\ }\textbf {\bibinfo {volume} {13}},\ \bibinfo {pages} {2433} (\bibinfo {year} {1976})}\BibitemShut {NoStop}%
\bibitem [{\citenamefont {Boman}\ \emph {et~al.}(1970)\citenamefont {Boman}, \citenamefont {Danielsen}, \citenamefont {Haaland}, \citenamefont {Jerslev}, \citenamefont {Schäffer}, \citenamefont {Sunde},\ and\ \citenamefont {Sørensen}}]{Boman1970}%
  \BibitemOpen
  \bibfield  {author} {\bibinfo {author} {\bibfnamefont {C.-E.}\ \bibnamefont {Boman}}, \bibinfo {author} {\bibfnamefont {J.}~\bibnamefont {Danielsen}}, \bibinfo {author} {\bibfnamefont {A.}~\bibnamefont {Haaland}}, \bibinfo {author} {\bibfnamefont {B.}~\bibnamefont {Jerslev}}, \bibinfo {author} {\bibfnamefont {C.~E.}\ \bibnamefont {Schäffer}}, \bibinfo {author} {\bibfnamefont {E.}~\bibnamefont {Sunde}},\ and\ \bibinfo {author} {\bibfnamefont {N.~A.}\ \bibnamefont {Sørensen}},\ }\bibfield  {title} {\bibinfo {title} {{Refinement of the Crystal Structure of Ruthenium Dioxide}},\ }\href {https://doi.org/10.3891/acta.chem.scand.24-0116} {\bibfield  {journal} {\bibinfo  {journal} {Acta Chemica Scandinavica}\ }\textbf {\bibinfo {volume} {24}},\ \bibinfo {pages} {116} (\bibinfo {year} {1970})}\BibitemShut {NoStop}%
\bibitem [{\citenamefont {Lee}\ \emph {et~al.}(2012)\citenamefont {Lee}, \citenamefont {Suntivich}, \citenamefont {May}, \citenamefont {Perry},\ and\ \citenamefont {Shao-Horn}}]{Lee2012}%
  \BibitemOpen
  \bibfield  {author} {\bibinfo {author} {\bibfnamefont {Y.}~\bibnamefont {Lee}}, \bibinfo {author} {\bibfnamefont {J.}~\bibnamefont {Suntivich}}, \bibinfo {author} {\bibfnamefont {K.~J.}\ \bibnamefont {May}}, \bibinfo {author} {\bibfnamefont {E.~E.}\ \bibnamefont {Perry}},\ and\ \bibinfo {author} {\bibfnamefont {Y.}~\bibnamefont {Shao-Horn}},\ }\bibfield  {title} {\bibinfo {title} {{Synthesis and activities of rutile IrO$_2$ and RuO$_2$ nanoparticles for oxygen evolution in acid and alkaline solutions}},\ }\href {https://doi.org/10.1021/jz2016507} {\bibfield  {journal} {\bibinfo  {journal} {Journal of Physical Chemistry Letters}\ }\textbf {\bibinfo {volume} {3}},\ \bibinfo {pages} {399} (\bibinfo {year} {2012})}\BibitemShut {NoStop}%
\bibitem [{\citenamefont {Ruf}\ \emph {et~al.}(2021)\citenamefont {Ruf}, \citenamefont {Paik}, \citenamefont {Schreiber}, \citenamefont {Nair}, \citenamefont {Miao}, \citenamefont {Kawasaki}, \citenamefont {Nelson}, \citenamefont {Faeth}, \citenamefont {Lee}, \citenamefont {Goodge}, \citenamefont {Pamuk}, \citenamefont {Fennie}, \citenamefont {Kourkoutis}, \citenamefont {Schlom},\ and\ \citenamefont {Shen}}]{Ruf2021}%
  \BibitemOpen
  \bibfield  {author} {\bibinfo {author} {\bibfnamefont {J.~P.}\ \bibnamefont {Ruf}}, \bibinfo {author} {\bibfnamefont {H.}~\bibnamefont {Paik}}, \bibinfo {author} {\bibfnamefont {N.~J.}\ \bibnamefont {Schreiber}}, \bibinfo {author} {\bibfnamefont {H.~P.}\ \bibnamefont {Nair}}, \bibinfo {author} {\bibfnamefont {L.}~\bibnamefont {Miao}}, \bibinfo {author} {\bibfnamefont {J.~K.}\ \bibnamefont {Kawasaki}}, \bibinfo {author} {\bibfnamefont {J.~N.}\ \bibnamefont {Nelson}}, \bibinfo {author} {\bibfnamefont {B.~D.}\ \bibnamefont {Faeth}}, \bibinfo {author} {\bibfnamefont {Y.}~\bibnamefont {Lee}}, \bibinfo {author} {\bibfnamefont {B.~H.}\ \bibnamefont {Goodge}}, \bibinfo {author} {\bibfnamefont {B.}~\bibnamefont {Pamuk}}, \bibinfo {author} {\bibfnamefont {C.~J.}\ \bibnamefont {Fennie}}, \bibinfo {author} {\bibfnamefont {L.~F.}\ \bibnamefont {Kourkoutis}}, \bibinfo {author} {\bibfnamefont {D.~G.}\ \bibnamefont {Schlom}},\ and\ \bibinfo {author} {\bibfnamefont {K.~M.}\ \bibnamefont {Shen}},\ }\bibfield  {title}
  {\bibinfo {title} {Strain-stabilized superconductivity},\ }\href {https://doi.org/10.1038/s41467-020-20252-7} {\bibfield  {journal} {\bibinfo  {journal} {Nature Communications}\ }\textbf {\bibinfo {volume} {12}},\ \bibinfo {pages} {1} (\bibinfo {year} {2021})}\BibitemShut {NoStop}%
\bibitem [{\citenamefont {Uchida}\ \emph {et~al.}(2020)\citenamefont {Uchida}, \citenamefont {Nomoto}, \citenamefont {Musashi}, \citenamefont {Arita},\ and\ \citenamefont {Kawasaki}}]{Uchida2020}%
  \BibitemOpen
  \bibfield  {author} {\bibinfo {author} {\bibfnamefont {M.}~\bibnamefont {Uchida}}, \bibinfo {author} {\bibfnamefont {T.}~\bibnamefont {Nomoto}}, \bibinfo {author} {\bibfnamefont {M.}~\bibnamefont {Musashi}}, \bibinfo {author} {\bibfnamefont {R.}~\bibnamefont {Arita}},\ and\ \bibinfo {author} {\bibfnamefont {M.}~\bibnamefont {Kawasaki}},\ }\bibfield  {title} {\bibinfo {title} {{Superconductivity in Uniquely Strained RuO$_2$ Films}},\ }\href {http://arxiv.org/abs/2008.12529} {\bibfield  {journal} {\bibinfo  {journal} {Physical Review Letters}\ }\textbf {\bibinfo {volume} {125}},\ \bibinfo {pages} {147001} (\bibinfo {year} {2020})}\BibitemShut {NoStop}%
\bibitem [{\citenamefont {Jovic}\ \emph {et~al.}(2018)\citenamefont {Jovic}, \citenamefont {Koch}, \citenamefont {Panda}, \citenamefont {Berger}, \citenamefont {Bugnon}, \citenamefont {Magrez}, \citenamefont {Smith}, \citenamefont {Biermann}, \citenamefont {Jozwiak}, \citenamefont {Bostwick}, \citenamefont {Rotenberg},\ and\ \citenamefont {Moser}}]{Jovic2018}%
  \BibitemOpen
  \bibfield  {author} {\bibinfo {author} {\bibfnamefont {V.}~\bibnamefont {Jovic}}, \bibinfo {author} {\bibfnamefont {R.~J.}\ \bibnamefont {Koch}}, \bibinfo {author} {\bibfnamefont {S.~K.}\ \bibnamefont {Panda}}, \bibinfo {author} {\bibfnamefont {H.}~\bibnamefont {Berger}}, \bibinfo {author} {\bibfnamefont {P.}~\bibnamefont {Bugnon}}, \bibinfo {author} {\bibfnamefont {A.}~\bibnamefont {Magrez}}, \bibinfo {author} {\bibfnamefont {K.~E.}\ \bibnamefont {Smith}}, \bibinfo {author} {\bibfnamefont {S.}~\bibnamefont {Biermann}}, \bibinfo {author} {\bibfnamefont {C.}~\bibnamefont {Jozwiak}}, \bibinfo {author} {\bibfnamefont {A.}~\bibnamefont {Bostwick}}, \bibinfo {author} {\bibfnamefont {E.}~\bibnamefont {Rotenberg}},\ and\ \bibinfo {author} {\bibfnamefont {S.}~\bibnamefont {Moser}},\ }\bibfield  {title} {\bibinfo {title} {{Dirac nodal lines and flat-band surface state in the functional oxide RuO$_2$}},\ }\href {https://doi.org/10.1103/PhysRevB.98.241101} {\bibfield  {journal} {\bibinfo  {journal} {Physical Review
  B}\ }\textbf {\bibinfo {volume} {98}},\ \bibinfo {pages} {241101} (\bibinfo {year} {2018})}\BibitemShut {NoStop}%
\bibitem [{\citenamefont {Ryden}\ and\ \citenamefont {Lawson}(1970)}]{Ryden1970}%
  \BibitemOpen
  \bibfield  {author} {\bibinfo {author} {\bibfnamefont {W.~D.}\ \bibnamefont {Ryden}}\ and\ \bibinfo {author} {\bibfnamefont {A.~W.}\ \bibnamefont {Lawson}},\ }\bibfield  {title} {\bibinfo {title} {{Magnetic susceptibility of IrO$_2$ and RuO$_2$}},\ }\href {https://doi.org/10.1063/1.1672908} {\bibfield  {journal} {\bibinfo  {journal} {The Journal of Chemical Physics}\ }\textbf {\bibinfo {volume} {52}},\ \bibinfo {pages} {6058} (\bibinfo {year} {1970})}\BibitemShut {NoStop}%
\bibitem [{\citenamefont {Passenheim}\ and\ \citenamefont {McCollum}(1969)}]{Passenheim1969}%
  \BibitemOpen
  \bibfield  {author} {\bibinfo {author} {\bibfnamefont {B.~C.}\ \bibnamefont {Passenheim}}\ and\ \bibinfo {author} {\bibfnamefont {D.~C.}\ \bibnamefont {McCollum}},\ }\bibfield  {title} {\bibinfo {title} {{Heat capacity of RuO$_2$ and IrO$_2$ between 0.54° and 10°K}},\ }\href {https://doi.org/10.1063/1.1671725} {\bibfield  {journal} {\bibinfo  {journal} {The Journal of Chemical Physics}\ }\textbf {\bibinfo {volume} {51}},\ \bibinfo {pages} {320} (\bibinfo {year} {1969})}\BibitemShut {NoStop}%
\bibitem [{\citenamefont {Glassford}\ and\ \citenamefont {Chelikowsky}(1994)}]{Glassford1994}%
  \BibitemOpen
  \bibfield  {author} {\bibinfo {author} {\bibfnamefont {K.~M.}\ \bibnamefont {Glassford}}\ and\ \bibinfo {author} {\bibfnamefont {J.~R.}\ \bibnamefont {Chelikowsky}},\ }\bibfield  {title} {\bibinfo {title} {{Electron transport properties in RuO$_2$ rutile}},\ }\href {https://doi.org/10.1103/PhysRevB.49.7107} {\bibfield  {journal} {\bibinfo  {journal} {Physical Review B}\ }\textbf {\bibinfo {volume} {49}},\ \bibinfo {pages} {7107} (\bibinfo {year} {1994})}\BibitemShut {NoStop}%
\bibitem [{\citenamefont {Zhu}\ \emph {et~al.}(2019)\citenamefont {Zhu}, \citenamefont {Strempfer}, \citenamefont {Rao}, \citenamefont {Occhialini}, \citenamefont {Pelliciari}, \citenamefont {Choi}, \citenamefont {Kawaguchi}, \citenamefont {You}, \citenamefont {Mitchell}, \citenamefont {Shao-Horn},\ and\ \citenamefont {Comin}}]{Zhu2019}%
  \BibitemOpen
  \bibfield  {author} {\bibinfo {author} {\bibfnamefont {Z.~H.}\ \bibnamefont {Zhu}}, \bibinfo {author} {\bibfnamefont {J.}~\bibnamefont {Strempfer}}, \bibinfo {author} {\bibfnamefont {R.~R.}\ \bibnamefont {Rao}}, \bibinfo {author} {\bibfnamefont {C.~A.}\ \bibnamefont {Occhialini}}, \bibinfo {author} {\bibfnamefont {J.}~\bibnamefont {Pelliciari}}, \bibinfo {author} {\bibfnamefont {Y.}~\bibnamefont {Choi}}, \bibinfo {author} {\bibfnamefont {T.}~\bibnamefont {Kawaguchi}}, \bibinfo {author} {\bibfnamefont {H.}~\bibnamefont {You}}, \bibinfo {author} {\bibfnamefont {J.~F.}\ \bibnamefont {Mitchell}}, \bibinfo {author} {\bibfnamefont {Y.}~\bibnamefont {Shao-Horn}},\ and\ \bibinfo {author} {\bibfnamefont {R.}~\bibnamefont {Comin}},\ }\bibfield  {title} {\bibinfo {title} {{Anomalous Antiferromagnetism in Metallic RuO$_2$ Determined by Resonant X-ray Scattering}},\ }\href {https://doi.org/10.1103/PhysRevLett.122.017202} {\bibfield  {journal} {\bibinfo  {journal} {Physical Review Letters}\ }\textbf {\bibinfo {volume}
  {122}},\ \bibinfo {pages} {017202} (\bibinfo {year} {2019})}\BibitemShut {NoStop}%
\bibitem [{\citenamefont {Keßler}\ \emph {et~al.}(2024)\citenamefont {Keßler}, \citenamefont {Garcia-Gassull}, \citenamefont {Suter}, \citenamefont {Prokscha}, \citenamefont {Salman}, \citenamefont {Khalyavin}, \citenamefont {Manuel}, \citenamefont {Orlandi}, \citenamefont {Mazin}, \citenamefont {Valenti},\ and\ \citenamefont {Moser}}]{Kesler2024}%
  \BibitemOpen
  \bibfield  {author} {\bibinfo {author} {\bibfnamefont {P.}~\bibnamefont {Keßler}}, \bibinfo {author} {\bibfnamefont {L.}~\bibnamefont {Garcia-Gassull}}, \bibinfo {author} {\bibfnamefont {A.}~\bibnamefont {Suter}}, \bibinfo {author} {\bibfnamefont {T.}~\bibnamefont {Prokscha}}, \bibinfo {author} {\bibfnamefont {Z.}~\bibnamefont {Salman}}, \bibinfo {author} {\bibfnamefont {D.}~\bibnamefont {Khalyavin}}, \bibinfo {author} {\bibfnamefont {P.}~\bibnamefont {Manuel}}, \bibinfo {author} {\bibfnamefont {F.}~\bibnamefont {Orlandi}}, \bibinfo {author} {\bibfnamefont {I.~I.}\ \bibnamefont {Mazin}}, \bibinfo {author} {\bibfnamefont {R.}~\bibnamefont {Valenti}},\ and\ \bibinfo {author} {\bibfnamefont {S.}~\bibnamefont {Moser}},\ }\bibfield  {title} {\bibinfo {title} {{Absence of magnetic order in RuO$_2$: insights from $\mu$SR spectroscopy and neutron diffraction}},\ }\href@noop {} {\bibfield  {journal} {\bibinfo  {journal} {npJ Spintronics}\ }\textbf {\bibinfo {volume} {2}},\ \bibinfo {pages} {50} (\bibinfo {year}
  {2024})}\BibitemShut {NoStop}%
\bibitem [{\citenamefont {Kiefer}\ \emph {et~al.}(2024)\citenamefont {Kiefer}, \citenamefont {Wirth}, \citenamefont {Bertin}, \citenamefont {Becker}, \citenamefont {Bohatý}, \citenamefont {Schmalzl}, \citenamefont {Stunault}, \citenamefont {Rodríguez-Velamazán}, \citenamefont {Fabelo},\ and\ \citenamefont {Braden}}]{Kiefer2024}%
  \BibitemOpen
  \bibfield  {author} {\bibinfo {author} {\bibfnamefont {L.}~\bibnamefont {Kiefer}}, \bibinfo {author} {\bibfnamefont {F.}~\bibnamefont {Wirth}}, \bibinfo {author} {\bibfnamefont {A.}~\bibnamefont {Bertin}}, \bibinfo {author} {\bibfnamefont {P.}~\bibnamefont {Becker}}, \bibinfo {author} {\bibfnamefont {L.}~\bibnamefont {Bohatý}}, \bibinfo {author} {\bibfnamefont {K.}~\bibnamefont {Schmalzl}}, \bibinfo {author} {\bibfnamefont {A.}~\bibnamefont {Stunault}}, \bibinfo {author} {\bibfnamefont {J.~A.}\ \bibnamefont {Rodríguez-Velamazán}}, \bibinfo {author} {\bibfnamefont {O.}~\bibnamefont {Fabelo}},\ and\ \bibinfo {author} {\bibfnamefont {M.}~\bibnamefont {Braden}},\ }\bibfield  {title} {\bibinfo {title} {{Crystal structure and absence of magnetic order in single crystalline RuO$_2$}},\ }\href {https://doi.org/10.1088/1361-648X/adad2a} {\bibfield  {journal} {\bibinfo  {journal} {Journal of Physics: Condensed Matter}\ }\textbf {\bibinfo {volume} {37}},\ \bibinfo {pages} {135801} (\bibinfo {year}
  {2024})}\BibitemShut {NoStop}%
\bibitem [{\citenamefont {Pawula}\ \emph {et~al.}(2024)\citenamefont {Pawula}, \citenamefont {Fakih}, \citenamefont {Daou}, \citenamefont {Hébert}, \citenamefont {Mordvinova}, \citenamefont {Lebedev}, \citenamefont {Pelloquin},\ and\ \citenamefont {Maignan}}]{Pawula2024}%
  \BibitemOpen
  \bibfield  {author} {\bibinfo {author} {\bibfnamefont {F.}~\bibnamefont {Pawula}}, \bibinfo {author} {\bibfnamefont {A.}~\bibnamefont {Fakih}}, \bibinfo {author} {\bibfnamefont {R.}~\bibnamefont {Daou}}, \bibinfo {author} {\bibfnamefont {S.}~\bibnamefont {Hébert}}, \bibinfo {author} {\bibfnamefont {N.}~\bibnamefont {Mordvinova}}, \bibinfo {author} {\bibfnamefont {O.}~\bibnamefont {Lebedev}}, \bibinfo {author} {\bibfnamefont {D.}~\bibnamefont {Pelloquin}},\ and\ \bibinfo {author} {\bibfnamefont {A.}~\bibnamefont {Maignan}},\ }\bibfield  {title} {\bibinfo {title} {{Multiband transport in RuO$_2$}},\ }\href {https://doi.org/10.1103/PhysRevB.110.064432} {\bibfield  {journal} {\bibinfo  {journal} {Physical Review B}\ }\textbf {\bibinfo {volume} {110}},\ \bibinfo {pages} {064432} (\bibinfo {year} {2024})}\BibitemShut {NoStop}%
\bibitem [{\citenamefont {Smolyanyuk}\ \emph {et~al.}(2023)\citenamefont {Smolyanyuk}, \citenamefont {Mazin}, \citenamefont {Garcia-Gassull},\ and\ \citenamefont {Valenti}}]{Smolyanyuk2023}%
  \BibitemOpen
  \bibfield  {author} {\bibinfo {author} {\bibfnamefont {A.}~\bibnamefont {Smolyanyuk}}, \bibinfo {author} {\bibfnamefont {I.~I.}\ \bibnamefont {Mazin}}, \bibinfo {author} {\bibfnamefont {L.}~\bibnamefont {Garcia-Gassull}},\ and\ \bibinfo {author} {\bibfnamefont {R.}~\bibnamefont {Valenti}},\ }\bibfield  {title} {\bibinfo {title} {{Fragility of the magnetic order in the prototypical altermagnet RuO$_2$}},\ }\href {https://doi.org/10.1103/PhysRevB.109.134424} {\bibfield  {journal} {\bibinfo  {journal} {Physical Review B}\ }\textbf {\bibinfo {volume} {109}},\ \bibinfo {pages} {134424} (\bibinfo {year} {2023})}\BibitemShut {NoStop}%
\bibitem [{\citenamefont {Wenzel}\ \emph {et~al.}(2024)\citenamefont {Wenzel}, \citenamefont {Uykur}, \citenamefont {Rößler}, \citenamefont {Schmidt}, \citenamefont {Janson}, \citenamefont {Tiwari}, \citenamefont {Dressel},\ and\ \citenamefont {Tsirlin}}]{Wenzel2024}%
  \BibitemOpen
  \bibfield  {author} {\bibinfo {author} {\bibfnamefont {M.}~\bibnamefont {Wenzel}}, \bibinfo {author} {\bibfnamefont {E.}~\bibnamefont {Uykur}}, \bibinfo {author} {\bibfnamefont {S.}~\bibnamefont {Rößler}}, \bibinfo {author} {\bibfnamefont {M.}~\bibnamefont {Schmidt}}, \bibinfo {author} {\bibfnamefont {O.}~\bibnamefont {Janson}}, \bibinfo {author} {\bibfnamefont {A.}~\bibnamefont {Tiwari}}, \bibinfo {author} {\bibfnamefont {M.}~\bibnamefont {Dressel}},\ and\ \bibinfo {author} {\bibfnamefont {A.~A.}\ \bibnamefont {Tsirlin}},\ }\bibfield  {title} {\bibinfo {title} {{Fermi-liquid behavior of non-altermagnetic RuO$_2$}},\ }\href@noop {} {\bibfield  {journal} {\bibinfo  {journal} {Physical Review B}\ }\textbf {\bibinfo {volume} {111}},\ \bibinfo {pages} {L041115} (\bibinfo {year} {2024})}\BibitemShut {NoStop}%
\bibitem [{\citenamefont {Yumnam}\ \emph {et~al.}(2025)\citenamefont {Yumnam}, \citenamefont {Raghuvanshi}, \citenamefont {Budai}, \citenamefont {Bansal}, \citenamefont {Bocklage}, \citenamefont {Abernathy}, \citenamefont {Cheng}, \citenamefont {Said}, \citenamefont {Mazin}, \citenamefont {Zhou}, \citenamefont {Frandsen}, \citenamefont {Parker}, \citenamefont {Lindsay}, \citenamefont {Cooper}, \citenamefont {Manley},\ and\ \citenamefont {Hermann}}]{Yumnam2025}%
  \BibitemOpen
  \bibfield  {author} {\bibinfo {author} {\bibfnamefont {G.}~\bibnamefont {Yumnam}}, \bibinfo {author} {\bibfnamefont {P.~R.}\ \bibnamefont {Raghuvanshi}}, \bibinfo {author} {\bibfnamefont {J.~D.}\ \bibnamefont {Budai}}, \bibinfo {author} {\bibfnamefont {D.}~\bibnamefont {Bansal}}, \bibinfo {author} {\bibfnamefont {L.}~\bibnamefont {Bocklage}}, \bibinfo {author} {\bibfnamefont {D.}~\bibnamefont {Abernathy}}, \bibinfo {author} {\bibfnamefont {Y.}~\bibnamefont {Cheng}}, \bibinfo {author} {\bibfnamefont {A.}~\bibnamefont {Said}}, \bibinfo {author} {\bibfnamefont {I.~I.}\ \bibnamefont {Mazin}}, \bibinfo {author} {\bibfnamefont {H.}~\bibnamefont {Zhou}}, \bibinfo {author} {\bibfnamefont {B.~A.}\ \bibnamefont {Frandsen}}, \bibinfo {author} {\bibfnamefont {D.~S.}\ \bibnamefont {Parker}}, \bibinfo {author} {\bibfnamefont {L.~R.}\ \bibnamefont {Lindsay}}, \bibinfo {author} {\bibfnamefont {V.~R.}\ \bibnamefont {Cooper}}, \bibinfo {author} {\bibfnamefont {M.~E.}\ \bibnamefont {Manley}},\ and\ \bibinfo {author}
  {\bibfnamefont {R.~P.}\ \bibnamefont {Hermann}},\ }\bibfield  {title} {\bibinfo {title} {{Constraints on magnetism and correlations in RuO$_2$ from lattice dynamics and M\"ossbauer spectroscopy}},\ }\href {http://arxiv.org/abs/2505.03250} {\bibfield  {journal} {\bibinfo  {journal} {arXiv}\ } (\bibinfo {year} {2025})}\BibitemShut {NoStop}%
\bibitem [{\citenamefont {Paul}\ \emph {et~al.}(2025)\citenamefont {Paul}, \citenamefont {Mattoni}, \citenamefont {Matsuki}, \citenamefont {Johnson}, \citenamefont {Sow}, \citenamefont {Yonezawa},\ and\ \citenamefont {Maeno}}]{Paul2025}%
  \BibitemOpen
  \bibfield  {author} {\bibinfo {author} {\bibfnamefont {S.}~\bibnamefont {Paul}}, \bibinfo {author} {\bibfnamefont {G.}~\bibnamefont {Mattoni}}, \bibinfo {author} {\bibfnamefont {H.}~\bibnamefont {Matsuki}}, \bibinfo {author} {\bibfnamefont {T.}~\bibnamefont {Johnson}}, \bibinfo {author} {\bibfnamefont {C.}~\bibnamefont {Sow}}, \bibinfo {author} {\bibfnamefont {S.}~\bibnamefont {Yonezawa}},\ and\ \bibinfo {author} {\bibfnamefont {Y.}~\bibnamefont {Maeno}},\ }\bibfield  {title} {\bibinfo {title} {{Growth of ultra-clean single crystals of RuO$_2$}},\ }\href {http://arxiv.org/abs/2505.07201} {\bibfield  {journal} {\bibinfo  {journal} {arXiv}\ } (\bibinfo {year} {2025})}\BibitemShut {NoStop}%
\bibitem [{\citenamefont {Liu}\ \emph {et~al.}(2024{\natexlab{b}})\citenamefont {Liu}, \citenamefont {Zhan}, \citenamefont {Li}, \citenamefont {Liu}, \citenamefont {Cheng}, \citenamefont {Shi}, \citenamefont {Deng}, \citenamefont {Zhang}, \citenamefont {Li}, \citenamefont {Ding}, \citenamefont {Jiang}, \citenamefont {Ye}, \citenamefont {Liu}, \citenamefont {Jiang}, \citenamefont {Wang}, \citenamefont {Li}, \citenamefont {Xie}, \citenamefont {Wang}, \citenamefont {Qiao}, \citenamefont {Wen}, \citenamefont {Sun},\ and\ \citenamefont {Shen}}]{Liu2024b}%
  \BibitemOpen
  \bibfield  {author} {\bibinfo {author} {\bibfnamefont {J.}~\bibnamefont {Liu}}, \bibinfo {author} {\bibfnamefont {J.}~\bibnamefont {Zhan}}, \bibinfo {author} {\bibfnamefont {T.}~\bibnamefont {Li}}, \bibinfo {author} {\bibfnamefont {J.}~\bibnamefont {Liu}}, \bibinfo {author} {\bibfnamefont {S.}~\bibnamefont {Cheng}}, \bibinfo {author} {\bibfnamefont {Y.}~\bibnamefont {Shi}}, \bibinfo {author} {\bibfnamefont {L.}~\bibnamefont {Deng}}, \bibinfo {author} {\bibfnamefont {M.}~\bibnamefont {Zhang}}, \bibinfo {author} {\bibfnamefont {C.}~\bibnamefont {Li}}, \bibinfo {author} {\bibfnamefont {J.}~\bibnamefont {Ding}}, \bibinfo {author} {\bibfnamefont {Q.}~\bibnamefont {Jiang}}, \bibinfo {author} {\bibfnamefont {M.}~\bibnamefont {Ye}}, \bibinfo {author} {\bibfnamefont {Z.}~\bibnamefont {Liu}}, \bibinfo {author} {\bibfnamefont {Z.}~\bibnamefont {Jiang}}, \bibinfo {author} {\bibfnamefont {S.}~\bibnamefont {Wang}}, \bibinfo {author} {\bibfnamefont {Q.}~\bibnamefont {Li}}, \bibinfo {author} {\bibfnamefont
  {Y.}~\bibnamefont {Xie}}, \bibinfo {author} {\bibfnamefont {Y.}~\bibnamefont {Wang}}, \bibinfo {author} {\bibfnamefont {S.}~\bibnamefont {Qiao}}, \bibinfo {author} {\bibfnamefont {J.}~\bibnamefont {Wen}}, \bibinfo {author} {\bibfnamefont {Y.}~\bibnamefont {Sun}},\ and\ \bibinfo {author} {\bibfnamefont {D.}~\bibnamefont {Shen}},\ }\bibfield  {title} {\bibinfo {title} {{Absence of altermagnetic spin splitting character in rutile oxide RuO$_2$}},\ }\href {https://doi.org/10.1103/PhysRevLett.133.176401} {\bibfield  {journal} {\bibinfo  {journal} {Physical Review Letters}\ }\textbf {\bibinfo {volume} {133}},\ \bibinfo {pages} {176401} (\bibinfo {year} {2024}{\natexlab{b}})}\BibitemShut {NoStop}%
\bibitem [{\citenamefont {Wang}\ \emph {et~al.}(2025)\citenamefont {Wang}, \citenamefont {Zhang}, \citenamefont {Tian}, \citenamefont {Yu},\ and\ \citenamefont {Kagawa}}]{Wang2025}%
  \BibitemOpen
  \bibfield  {author} {\bibinfo {author} {\bibfnamefont {M.}~\bibnamefont {Wang}}, \bibinfo {author} {\bibfnamefont {J.}~\bibnamefont {Zhang}}, \bibinfo {author} {\bibfnamefont {D.}~\bibnamefont {Tian}}, \bibinfo {author} {\bibfnamefont {P.}~\bibnamefont {Yu}},\ and\ \bibinfo {author} {\bibfnamefont {F.}~\bibnamefont {Kagawa}},\ }\bibfield  {title} {\bibinfo {title} {{Unveiling an in-plane Hall effect in rutile RuO$_2$ films}},\ }\href {https://doi.org/10.1038/s42005-025-01943-3} {\bibfield  {journal} {\bibinfo  {journal} {Communications Physics}\ }\textbf {\bibinfo {volume} {8}},\ \bibinfo {pages} {28} (\bibinfo {year} {2025})}\BibitemShut {NoStop}%
\bibitem [{\citenamefont {Dmitrienko}(1983)}]{Dmitrienko1983}%
  \BibitemOpen
  \bibfield  {author} {\bibinfo {author} {\bibfnamefont {V.~E.}\ \bibnamefont {Dmitrienko}},\ }\bibfield  {title} {\bibinfo {title} {Forbidden reflections due to anisotropic x-ray susceptibility of crystals},\ }\href@noop {} {\bibfield  {journal} {\bibinfo  {journal} {Acta Crystallographica}\ }\textbf {\bibinfo {volume} {39}},\ \bibinfo {pages} {29} (\bibinfo {year} {1983})}\BibitemShut {NoStop}%
\bibitem [{\citenamefont {Templeton}\ and\ \citenamefont {Templeton}(1980)}]{Templeton1980}%
  \BibitemOpen
  \bibfield  {author} {\bibinfo {author} {\bibfnamefont {D.~H.}\ \bibnamefont {Templeton}}\ and\ \bibinfo {author} {\bibfnamefont {L.~K.}\ \bibnamefont {Templeton}},\ }\bibfield  {title} {\bibinfo {title} {Polarized x-ray absorption and double refraction in vanadyl bisacetylacetonate},\ }\href@noop {} {\bibfield  {journal} {\bibinfo  {journal} {Act Crystallographica}\ }\textbf {\bibinfo {volume} {A36}},\ \bibinfo {pages} {237} (\bibinfo {year} {1980})}\BibitemShut {NoStop}%
\bibitem [{\citenamefont {Templeton}\ and\ \citenamefont {Templeton}(1982)}]{Templeton1982}%
  \BibitemOpen
  \bibfield  {author} {\bibinfo {author} {\bibfnamefont {D.~H.}\ \bibnamefont {Templeton}}\ and\ \bibinfo {author} {\bibfnamefont {L.~K.}\ \bibnamefont {Templeton}},\ }\bibfield  {title} {\bibinfo {title} {X-ray dichroism and polarized anomalous scattering of the uranyl ion},\ }\href@noop {} {\bibfield  {journal} {\bibinfo  {journal} {Acta Crystallographica}\ }\textbf {\bibinfo {volume} {A38}},\ \bibinfo {pages} {62} (\bibinfo {year} {1982})}\BibitemShut {NoStop}%
\bibitem [{\citenamefont {Occhialini}\ \emph {et~al.}(2022)\citenamefont {Occhialini}, \citenamefont {Martins}, \citenamefont {Fan}, \citenamefont {Bisogni}, \citenamefont {Yasunami}, \citenamefont {Musashi}, \citenamefont {Kawasaki}, \citenamefont {Uchida}, \citenamefont {Comin},\ and\ \citenamefont {Pelliciari}}]{Occhialini2022}%
  \BibitemOpen
  \bibfield  {author} {\bibinfo {author} {\bibfnamefont {C.~A.}\ \bibnamefont {Occhialini}}, \bibinfo {author} {\bibfnamefont {L.~G.}\ \bibnamefont {Martins}}, \bibinfo {author} {\bibfnamefont {S.}~\bibnamefont {Fan}}, \bibinfo {author} {\bibfnamefont {V.}~\bibnamefont {Bisogni}}, \bibinfo {author} {\bibfnamefont {T.}~\bibnamefont {Yasunami}}, \bibinfo {author} {\bibfnamefont {M.}~\bibnamefont {Musashi}}, \bibinfo {author} {\bibfnamefont {M.}~\bibnamefont {Kawasaki}}, \bibinfo {author} {\bibfnamefont {M.}~\bibnamefont {Uchida}}, \bibinfo {author} {\bibfnamefont {R.}~\bibnamefont {Comin}},\ and\ \bibinfo {author} {\bibfnamefont {J.}~\bibnamefont {Pelliciari}},\ }\bibfield  {title} {\bibinfo {title} {Strain-modulated anisotropic electronic structure in superconducting {RuO$_2$} films},\ }\href@noop {} {\bibfield  {journal} {\bibinfo  {journal} {Physical Review Materials}\ }\textbf {\bibinfo {volume} {6}},\ \bibinfo {pages} {084802} (\bibinfo {year} {2022})}\BibitemShut {NoStop}%
\bibitem [{\citenamefont {Gregory}\ \emph {et~al.}(2022)\citenamefont {Gregory}, \citenamefont {Strempfer}, \citenamefont {Weinstock}, \citenamefont {Ruf}, \citenamefont {Sun}, \citenamefont {Nair}, \citenamefont {Schreiber}, \citenamefont {Schlom}, \citenamefont {Shen},\ and\ \citenamefont {Singer}}]{Gregory2022}%
  \BibitemOpen
  \bibfield  {author} {\bibinfo {author} {\bibfnamefont {B.~Z.}\ \bibnamefont {Gregory}}, \bibinfo {author} {\bibfnamefont {J.}~\bibnamefont {Strempfer}}, \bibinfo {author} {\bibfnamefont {D.}~\bibnamefont {Weinstock}}, \bibinfo {author} {\bibfnamefont {J.~P.}\ \bibnamefont {Ruf}}, \bibinfo {author} {\bibfnamefont {Y.}~\bibnamefont {Sun}}, \bibinfo {author} {\bibfnamefont {H.}~\bibnamefont {Nair}}, \bibinfo {author} {\bibfnamefont {N.~J.}\ \bibnamefont {Schreiber}}, \bibinfo {author} {\bibfnamefont {D.~G.}\ \bibnamefont {Schlom}}, \bibinfo {author} {\bibfnamefont {K.~M.}\ \bibnamefont {Shen}},\ and\ \bibinfo {author} {\bibfnamefont {A.}~\bibnamefont {Singer}},\ }\bibfield  {title} {\bibinfo {title} {{Strain-induced orbital-energy shift in antiferromagnetic RuO$_2$ revealed by resonant elastic x-ray scattering}},\ }\href {https://doi.org/10.1103/PhysRevB.106.195135} {\bibfield  {journal} {\bibinfo  {journal} {Physical Review B}\ }\textbf {\bibinfo {volume} {106}},\ \bibinfo {pages} {195135} (\bibinfo {year}
  {2022})}\BibitemShut {NoStop}%
\bibitem [{\citenamefont {Lovesey}\ \emph {et~al.}(2022)\citenamefont {Lovesey}, \citenamefont {Khalyavin},\ and\ \citenamefont {van~der Laan}}]{Lovesey2022}%
  \BibitemOpen
  \bibfield  {author} {\bibinfo {author} {\bibfnamefont {S.~W.}\ \bibnamefont {Lovesey}}, \bibinfo {author} {\bibfnamefont {D.~D.}\ \bibnamefont {Khalyavin}},\ and\ \bibinfo {author} {\bibfnamefont {G.}~\bibnamefont {van~der Laan}},\ }\bibfield  {title} {\bibinfo {title} {{Magnetic properties of RuO$_2$ and charge-magnetic interference in Bragg diffraction of circularly polarized x-rays}},\ }\href {https://doi.org/10.1103/PhysRevB.105.014403} {\bibfield  {journal} {\bibinfo  {journal} {Physical Review B}\ }\textbf {\bibinfo {volume} {105}},\ \bibinfo {pages} {014403} (\bibinfo {year} {2022})}\BibitemShut {NoStop}%
\bibitem [{\citenamefont {Matteo}(2012)}]{Matteo2012}%
  \BibitemOpen
  \bibfield  {author} {\bibinfo {author} {\bibfnamefont {S.~D.}\ \bibnamefont {Matteo}},\ }\bibfield  {title} {\bibinfo {title} {{Resonant x-ray diffraction: Multipole interpretation}},\ }\href {https://doi.org/10.1088/0022-3727/45/16/163001} {\bibfield  {journal} {\bibinfo  {journal} {Journal of Physics D: Applied Physics}\ }\textbf {\bibinfo {volume} {45}},\ \bibinfo {pages} {163001} (\bibinfo {year} {2012})}\BibitemShut {NoStop}%
\bibitem [{sup()}]{suppref}%
  \BibitemOpen
  \href@noop {} {}\bibinfo {note} {See Supplemental Material at [URL will be inserted by publisher] for supporting data and details of the FDMNES calculations.}\BibitemShut {Stop}%
\bibitem [{\citenamefont {Occhialini}\ \emph {et~al.}(2021)\citenamefont {Occhialini}, \citenamefont {Bisogni}, \citenamefont {You}, \citenamefont {Barbour}, \citenamefont {Jarrige}, \citenamefont {Mitchell}, \citenamefont {Comin},\ and\ \citenamefont {Pelliciari}}]{Occhialini2021}%
  \BibitemOpen
  \bibfield  {author} {\bibinfo {author} {\bibfnamefont {C.~A.}\ \bibnamefont {Occhialini}}, \bibinfo {author} {\bibfnamefont {V.}~\bibnamefont {Bisogni}}, \bibinfo {author} {\bibfnamefont {H.}~\bibnamefont {You}}, \bibinfo {author} {\bibfnamefont {A.}~\bibnamefont {Barbour}}, \bibinfo {author} {\bibfnamefont {I.}~\bibnamefont {Jarrige}}, \bibinfo {author} {\bibfnamefont {J.~F.}\ \bibnamefont {Mitchell}}, \bibinfo {author} {\bibfnamefont {R.}~\bibnamefont {Comin}},\ and\ \bibinfo {author} {\bibfnamefont {J.}~\bibnamefont {Pelliciari}},\ }\bibfield  {title} {\bibinfo {title} {{Local electronic structure of rutile RuO$_2$}},\ }\href {https://doi.org/10.1103/PhysRevResearch.3.033214} {\bibfield  {journal} {\bibinfo  {journal} {Physical Review Research}\ }\textbf {\bibinfo {volume} {3}},\ \bibinfo {pages} {033214} (\bibinfo {year} {2021})}\BibitemShut {NoStop}%
\bibitem [{\citenamefont {Hirata}\ \emph {et~al.}(2013)\citenamefont {Hirata}, \citenamefont {Ohgushi}, \citenamefont {Yamaura}, \citenamefont {Ohsumi}, \citenamefont {Takeshita}, \citenamefont {Takata},\ and\ \citenamefont {Arima}}]{Hirata2013}%
  \BibitemOpen
  \bibfield  {author} {\bibinfo {author} {\bibfnamefont {Y.}~\bibnamefont {Hirata}}, \bibinfo {author} {\bibfnamefont {K.}~\bibnamefont {Ohgushi}}, \bibinfo {author} {\bibfnamefont {J.~I.}\ \bibnamefont {Yamaura}}, \bibinfo {author} {\bibfnamefont {H.}~\bibnamefont {Ohsumi}}, \bibinfo {author} {\bibfnamefont {S.}~\bibnamefont {Takeshita}}, \bibinfo {author} {\bibfnamefont {M.}~\bibnamefont {Takata}},\ and\ \bibinfo {author} {\bibfnamefont {T.~H.}\ \bibnamefont {Arima}},\ }\bibfield  {title} {\bibinfo {title} {{Complex orbital state stabilized by strong spin-orbit coupling in a metallic iridium oxide IrO$_2$}},\ }\href {https://doi.org/10.1103/PhysRevB.87.161111} {\bibfield  {journal} {\bibinfo  {journal} {Physical Review B}\ }\textbf {\bibinfo {volume} {87}},\ \bibinfo {pages} {161111} (\bibinfo {year} {2013})}\BibitemShut {NoStop}%
\bibitem [{\citenamefont {Kirfel}\ \emph {et~al.}(1991)\citenamefont {Kirfel}, \citenamefont {Petcov},\ and\ \citenamefont {Eichhorn}}]{Kirfel1991}%
  \BibitemOpen
  \bibfield  {author} {\bibinfo {author} {\bibfnamefont {A.}~\bibnamefont {Kirfel}}, \bibinfo {author} {\bibfnamefont {A.}~\bibnamefont {Petcov}},\ and\ \bibinfo {author} {\bibfnamefont {K.}~\bibnamefont {Eichhorn}},\ }\bibfield  {title} {\bibinfo {title} {Anisotropy of anomalous dispersion in x-ray diffraction},\ }\href@noop {} {\bibfield  {journal} {\bibinfo  {journal} {Acta Crystallographica}\ }\textbf {\bibinfo {volume} {A47}},\ \bibinfo {pages} {180} (\bibinfo {year} {1991})}\BibitemShut {NoStop}%
\bibitem [{\citenamefont {Sawai}\ \emph {et~al.}(2003)\citenamefont {Sawai}, \citenamefont {Kokubun},\ and\ \citenamefont {Ishida}}]{Sawai2003}%
  \BibitemOpen
  \bibfield  {author} {\bibinfo {author} {\bibfnamefont {H.}~\bibnamefont {Sawai}}, \bibinfo {author} {\bibfnamefont {J.}~\bibnamefont {Kokubun}},\ and\ \bibinfo {author} {\bibfnamefont {K.}~\bibnamefont {Ishida}},\ }\bibfield  {title} {\bibinfo {title} {{Anisotropic resonant x-ray scattering in rutile, TiO$_2$}},\ }\href@noop {} {\bibfield  {journal} {\bibinfo  {journal} {Materials Science Photon Factory Activity Report}\ }\textbf {\bibinfo {volume} {20 Part B}},\ \bibinfo {pages} {122} (\bibinfo {year} {2003})}\BibitemShut {NoStop}%
\bibitem [{Note1()}]{Note1}%
  \BibitemOpen
  \bibinfo {note} {In addition, a time-odd charge-magnetic interference term $\propto \cos (2\psi )$ occurs in the $|F_{\sigma \pi '}|^2$ channel which would be observable in single AFM domain samples. We neglect this term here as it does not affect the main conclusions. Further details are provided in the SM \cite {suppref}.}\BibitemShut {Stop}%
\bibitem [{\citenamefont {Zegkinoglou}\ \emph {et~al.}(2005)\citenamefont {Zegkinoglou}, \citenamefont {Strempfer}, \citenamefont {Nelson}, \citenamefont {Hill}, \citenamefont {Chakhalian}, \citenamefont {Bernhard}, \citenamefont {Lang}, \citenamefont {Srajer}, \citenamefont {Fukazawa}, \citenamefont {Nakatsuji}, \citenamefont {Maeno},\ and\ \citenamefont {Keimer}}]{Zegkinoglou2005}%
  \BibitemOpen
  \bibfield  {author} {\bibinfo {author} {\bibfnamefont {I.}~\bibnamefont {Zegkinoglou}}, \bibinfo {author} {\bibfnamefont {J.}~\bibnamefont {Strempfer}}, \bibinfo {author} {\bibfnamefont {C.~S.}\ \bibnamefont {Nelson}}, \bibinfo {author} {\bibfnamefont {J.~P.}\ \bibnamefont {Hill}}, \bibinfo {author} {\bibfnamefont {J.}~\bibnamefont {Chakhalian}}, \bibinfo {author} {\bibfnamefont {C.}~\bibnamefont {Bernhard}}, \bibinfo {author} {\bibfnamefont {J.~C.}\ \bibnamefont {Lang}}, \bibinfo {author} {\bibfnamefont {G.}~\bibnamefont {Srajer}}, \bibinfo {author} {\bibfnamefont {H.}~\bibnamefont {Fukazawa}}, \bibinfo {author} {\bibfnamefont {S.}~\bibnamefont {Nakatsuji}}, \bibinfo {author} {\bibfnamefont {Y.}~\bibnamefont {Maeno}},\ and\ \bibinfo {author} {\bibfnamefont {B.}~\bibnamefont {Keimer}},\ }\bibfield  {title} {\bibinfo {title} {{Orbital ordering transition in Ca$_2$RuO$_4$ observed with resonant X-ray diffraction}},\ }\href@noop {} {\bibfield  {journal} {\bibinfo  {journal} {Physical Review Letters}\ }\textbf
  {\bibinfo {volume} {95}} (\bibinfo {year} {2005})}\BibitemShut {NoStop}%
\bibitem [{\citenamefont {Porter}\ \emph {et~al.}(2018)\citenamefont {Porter}, \citenamefont {Granata}, \citenamefont {Forte}, \citenamefont {Matteo}, \citenamefont {Cuoco}, \citenamefont {Fittipaldi}, \citenamefont {Vecchione},\ and\ \citenamefont {Bombardi}}]{Porter2018}%
  \BibitemOpen
  \bibfield  {author} {\bibinfo {author} {\bibfnamefont {D.~G.}\ \bibnamefont {Porter}}, \bibinfo {author} {\bibfnamefont {V.}~\bibnamefont {Granata}}, \bibinfo {author} {\bibfnamefont {F.}~\bibnamefont {Forte}}, \bibinfo {author} {\bibfnamefont {S.~D.}\ \bibnamefont {Matteo}}, \bibinfo {author} {\bibfnamefont {M.}~\bibnamefont {Cuoco}}, \bibinfo {author} {\bibfnamefont {R.}~\bibnamefont {Fittipaldi}}, \bibinfo {author} {\bibfnamefont {A.}~\bibnamefont {Vecchione}},\ and\ \bibinfo {author} {\bibfnamefont {A.}~\bibnamefont {Bombardi}},\ }\bibfield  {title} {\bibinfo {title} {{Magnetic anisotropy and orbital ordering in Ca$_2$RuO$_4$}},\ }\href@noop {} {\bibfield  {journal} {\bibinfo  {journal} {Physical Review B}\ }\textbf {\bibinfo {volume} {98}},\ \bibinfo {pages} {125142} (\bibinfo {year} {2018})}\BibitemShut {NoStop}%
\bibitem [{\citenamefont {Bohnenbuck}\ \emph {et~al.}(2008)\citenamefont {Bohnenbuck}, \citenamefont {Zegkinoglou}, \citenamefont {Strempfer}, \citenamefont {Schüßler-Langeheine}, \citenamefont {Nelson}, \citenamefont {Leininger}, \citenamefont {Wu}, \citenamefont {Schierle}, \citenamefont {Lang}, \citenamefont {Srajer}, \citenamefont {Ikeda}, \citenamefont {Yoshida}, \citenamefont {Iwata}, \citenamefont {Katano}, \citenamefont {Kikugawa},\ and\ \citenamefont {Keimer}}]{Bohnenbuck2008}%
  \BibitemOpen
  \bibfield  {author} {\bibinfo {author} {\bibfnamefont {B.}~\bibnamefont {Bohnenbuck}}, \bibinfo {author} {\bibfnamefont {I.}~\bibnamefont {Zegkinoglou}}, \bibinfo {author} {\bibfnamefont {J.}~\bibnamefont {Strempfer}}, \bibinfo {author} {\bibfnamefont {C.}~\bibnamefont {Schüßler-Langeheine}}, \bibinfo {author} {\bibfnamefont {C.~S.}\ \bibnamefont {Nelson}}, \bibinfo {author} {\bibfnamefont {P.}~\bibnamefont {Leininger}}, \bibinfo {author} {\bibfnamefont {H.~H.}\ \bibnamefont {Wu}}, \bibinfo {author} {\bibfnamefont {E.}~\bibnamefont {Schierle}}, \bibinfo {author} {\bibfnamefont {J.~C.}\ \bibnamefont {Lang}}, \bibinfo {author} {\bibfnamefont {G.}~\bibnamefont {Srajer}}, \bibinfo {author} {\bibfnamefont {S.~I.}\ \bibnamefont {Ikeda}}, \bibinfo {author} {\bibfnamefont {Y.}~\bibnamefont {Yoshida}}, \bibinfo {author} {\bibfnamefont {K.}~\bibnamefont {Iwata}}, \bibinfo {author} {\bibfnamefont {S.}~\bibnamefont {Katano}}, \bibinfo {author} {\bibfnamefont {N.}~\bibnamefont {Kikugawa}},\ and\ \bibinfo {author}
  {\bibfnamefont {B.}~\bibnamefont {Keimer}},\ }\bibfield  {title} {\bibinfo {title} {{Magnetic structure and orbital state of Ca$_3$Ru$_2$O$_7$ investigated by resonant x-ray diffraction}},\ }\href@noop {} {\bibfield  {journal} {\bibinfo  {journal} {Physical Review B}\ }\textbf {\bibinfo {volume} {77}},\ \bibinfo {pages} {224412} (\bibinfo {year} {2008})}\BibitemShut {NoStop}%
\bibitem [{\citenamefont {Bunau}\ and\ \citenamefont {Joly}(2009)}]{Bunau2009}%
  \BibitemOpen
  \bibfield  {author} {\bibinfo {author} {\bibfnamefont {O.}~\bibnamefont {Bunau}}\ and\ \bibinfo {author} {\bibfnamefont {Y.}~\bibnamefont {Joly}},\ }\bibfield  {title} {\bibinfo {title} {Self-consistent aspects of x-ray absorption calculations},\ }\href {https://doi.org/10.1088/0953-8984/21/34/345501} {\bibfield  {journal} {\bibinfo  {journal} {Journal of Physics: Condensed Matter}\ }\textbf {\bibinfo {volume} {21}},\ \bibinfo {pages} {345501} (\bibinfo {year} {2009})}\BibitemShut {NoStop}%
\bibitem [{\citenamefont {Hu}\ \emph {et~al.}(2000)\citenamefont {Hu}, \citenamefont {von Lips}, \citenamefont {Golden}, \citenamefont {Fink},\ and\ \citenamefont {Kaindl}}]{Hu2000}%
  \BibitemOpen
  \bibfield  {author} {\bibinfo {author} {\bibfnamefont {Z.}~\bibnamefont {Hu}}, \bibinfo {author} {\bibfnamefont {H.}~\bibnamefont {von Lips}}, \bibinfo {author} {\bibfnamefont {M.}~\bibnamefont {Golden}}, \bibinfo {author} {\bibfnamefont {J.}~\bibnamefont {Fink}},\ and\ \bibinfo {author} {\bibfnamefont {G.}~\bibnamefont {Kaindl}},\ }\bibfield  {title} {\bibinfo {title} {{Multiplet effects in the Ru x-ray-absorption spectra of Ru(IV) and Ru(V) compounds}},\ }\href {https://doi.org/10.1103/PhysRevB.61.5262} {\bibfield  {journal} {\bibinfo  {journal} {Physical Review B}\ }\textbf {\bibinfo {volume} {61}},\ \bibinfo {pages} {5262} (\bibinfo {year} {2000})}\BibitemShut {NoStop}%
\bibitem [{\citenamefont {Yuan}\ and\ \citenamefont {Zunger}(2023)}]{Yuan2023}%
  \BibitemOpen
  \bibfield  {author} {\bibinfo {author} {\bibfnamefont {L.~D.}\ \bibnamefont {Yuan}}\ and\ \bibinfo {author} {\bibfnamefont {A.}~\bibnamefont {Zunger}},\ }\bibfield  {title} {\bibinfo {title} {{Degeneracy Removal of Spin Bands in Collinear Antiferromagnets with Non-Interconvertible Spin-Structure Motif Pair}},\ }\href@noop {} {\bibfield  {journal} {\bibinfo  {journal} {Advanced Materials}\ }\textbf {\bibinfo {volume} {35}},\ \bibinfo {pages} {2211966} (\bibinfo {year} {2023})}\BibitemShut {NoStop}%
\bibitem [{\citenamefont {Bergevin}\ and\ \citenamefont {Brunel}(1972)}]{Bergevin1972}%
  \BibitemOpen
  \bibfield  {author} {\bibinfo {author} {\bibfnamefont {F.~D.}\ \bibnamefont {Bergevin}}\ and\ \bibinfo {author} {\bibfnamefont {M.}~\bibnamefont {Brunel}},\ }\bibfield  {title} {\bibinfo {title} {{Observation of magnetic superlattice peaks by X-ray diffraction on an antiferromagnetic NiO crystal}},\ }\href@noop {} {\bibfield  {journal} {\bibinfo  {journal} {Physics Letters}\ }\textbf {\bibinfo {volume} {39A}},\ \bibinfo {pages} {141} (\bibinfo {year} {1972})}\BibitemShut {NoStop}%
\bibitem [{\citenamefont {Strempfer}\ \emph {et~al.}(1996)\citenamefont {Strempfer}, \citenamefont {Bruckel}, \citenamefont {Rott}, \citenamefont {Schneider}, \citenamefont {Liss},\ and\ \citenamefont {Tscrmntscher}}]{Strempfer1996}%
  \BibitemOpen
  \bibfield  {author} {\bibinfo {author} {\bibfnamefont {J.}~\bibnamefont {Strempfer}}, \bibinfo {author} {\bibfnamefont {T.}~\bibnamefont {Bruckel}}, \bibinfo {author} {\bibfnamefont {U.}~\bibnamefont {Rott}}, \bibinfo {author} {\bibfnamefont {J.~R.}\ \bibnamefont {Schneider}}, \bibinfo {author} {\bibfnamefont {K.-D.}\ \bibnamefont {Liss}},\ and\ \bibinfo {author} {\bibfnamefont {T.}~\bibnamefont {Tscrmntscher}},\ }\bibfield  {title} {\bibinfo {title} {{The Non-Resonant Magnetic X-ray Scattering Cross Section of MnF$_2$ 2. High-Energy X-ray Diffraction at 80 keV}},\ }\href@noop {} {\bibfield  {journal} {\bibinfo  {journal} {Acta Crystallographica}\ }\textbf {\bibinfo {volume} {52}},\ \bibinfo {pages} {438} (\bibinfo {year} {1996})}\BibitemShut {NoStop}%
\bibitem [{\citenamefont {Huang}\ \emph {et~al.}(1982)\citenamefont {Huang}, \citenamefont {Park},\ and\ \citenamefont {Pollak}}]{Huang1982}%
  \BibitemOpen
  \bibfield  {author} {\bibinfo {author} {\bibfnamefont {S.}~\bibnamefont {Huang}}, \bibinfo {author} {\bibfnamefont {H.~L.}\ \bibnamefont {Park}},\ and\ \bibinfo {author} {\bibfnamefont {F.~H.}\ \bibnamefont {Pollak}},\ }\bibfield  {title} {\bibinfo {title} {{Growth and characterization of RuO$_2$ single crystals}},\ }\href@noop {} {\bibfield  {journal} {\bibinfo  {journal} {Mat. Res. Bull.}\ }\textbf {\bibinfo {volume} {17}},\ \bibinfo {pages} {1305} (\bibinfo {year} {1982})}\BibitemShut {NoStop}%
\bibitem [{\citenamefont {Bolzan}\ \emph {et~al.}(1997)\citenamefont {Bolzan}, \citenamefont {Fong}, \citenamefont {Kennedy},\ and\ \citenamefont {Howard}}]{Bolzan1997}%
  \BibitemOpen
  \bibfield  {author} {\bibinfo {author} {\bibfnamefont {A.~A.}\ \bibnamefont {Bolzan}}, \bibinfo {author} {\bibfnamefont {C.}~\bibnamefont {Fong}}, \bibinfo {author} {\bibfnamefont {B.~J.}\ \bibnamefont {Kennedy}},\ and\ \bibinfo {author} {\bibfnamefont {C.~J.}\ \bibnamefont {Howard}},\ }\bibfield  {title} {\bibinfo {title} {Structural studies of rutile-type metal dioxides},\ }\href {https://doi.org/10.1107/S0108768197001468} {\bibfield  {journal} {\bibinfo  {journal} {Acta Crystallographica Section B: Structural Science}\ }\textbf {\bibinfo {volume} {53}},\ \bibinfo {pages} {373} (\bibinfo {year} {1997})}\BibitemShut {NoStop}%
\bibitem [{\citenamefont {Joly}\ \emph {et~al.}(2012)\citenamefont {Joly}, \citenamefont {Collins}, \citenamefont {Grenier}, \citenamefont {Tolentino},\ and\ \citenamefont {Santis}}]{Joly2012}%
  \BibitemOpen
  \bibfield  {author} {\bibinfo {author} {\bibfnamefont {Y.}~\bibnamefont {Joly}}, \bibinfo {author} {\bibfnamefont {S.~P.}\ \bibnamefont {Collins}}, \bibinfo {author} {\bibfnamefont {S.}~\bibnamefont {Grenier}}, \bibinfo {author} {\bibfnamefont {H.~C.}\ \bibnamefont {Tolentino}},\ and\ \bibinfo {author} {\bibfnamefont {M.~D.}\ \bibnamefont {Santis}},\ }\bibfield  {title} {\bibinfo {title} {Birefringence and polarization rotation in resonant x-ray diffraction},\ }\href@noop {} {\bibfield  {journal} {\bibinfo  {journal} {Physical Review B}\ }\textbf {\bibinfo {volume} {86}},\ \bibinfo {pages} {220101} (\bibinfo {year} {2012})}\BibitemShut {NoStop}%
\end{thebibliography}

%

\onecolumngrid
\section*{Appendix}

\twocolumngrid
{\it Appendix A: Samples} --- Bulk single crystals were grown in a tube furnace using the oxidative sublimation/crystallization method of Pollak et al. \cite{Huang1982}. The RuO$_2$ (99.95\%, electronic grade, Premion, Alfa Aesar, Ward Hill, MA) starting material was heated in an oxygen stream at the upstream end of the furnace at $\sim$1350 $^\circ$C and crystallized in a sharp temperature gradient at the downstream end at $\sim$1150 $^\circ$C. The resulting crystals ranged typically smaller than 5 mm in size with multiple facets including (110) and (100). $\mathbf{Q} = (100)$ measurements were performed on an as-grown (100) facet. $\mathbf{Q} = (001)$ measurements were performed on a cut and polished (001) facet.

Thin film measurements were performed on a 18-nm thick (001) RuO$_2$/TiO$_2$ epitaxial film grown by molecular beam epitaxy as described in Refs. \cite{Uchida2020, Occhialini2022}. The film is coherently strained resulting in a 2.2\% expansion of the in-plane $a$ and $b$ lattice constants with respect to bulk RuO$_2$. Further sample characterization is provided in Ref. \cite{Uchida2020}.\\

{\it Appendix B: Experimental Methods} --- Resonant X-ray diffraction (RXD) measurements at the Ru $L_3$-edge were performed at the 4-ID (ISR) beamline at the National Synchrotron Light Source II (NSLS-II), Brookhaven National Laboratory and at the I16 beamline at the Diamond Light Source (DLS). We utilized incident $\sigma$ polarization and analyzed the diffracted linear polarization ($\sigma'$/$\pi'$) using the (002) reflection of a graphite analyzer. The incident energy bandwidth provided by the Si (111) double crystal monochromator (NSLS-II) or Si (111) channel-cut monochromator (DLS) was $\Delta E_I \sim0.5$ eV. Silicon mirrors were used for harmonic rejection. Data at 4-ID were measured with a silicon drift detector (Amptek). Data at I16 were measured using a Merlin4x (Pilatus) area detector for polarization (un)resolved measurements. Variable temperature measurements ($T \sim$ 6 - 340 K) were performed using a closed cycle optical cryostat (Advanced Research Systems).  
\\

{\it Appendix C: FDMNES calculations} --- We performed Ru $L_3$-edge X-ray absorption spectroscopy (XAS) and resonant diffraction calculations using FDMNES \cite{Bunau2009}. Non-magnetic calculations proceed using a space group $P4_2/mnm$ and experimentally-determined lattice parameters $a = b = 4.49680$ \AA, $c = 3.10490$ \AA, using the oxygen position $w = 0.30530$ \cite{Bolzan1997}. For magnetic calculations, we assumed the $P4_2'/mnm'$ magnetic space group with local spin axis along the $c$ axis with a $\mathbf{k} = 0$ AFM ordering, using the same lattice parameters as for the non-magnetic calculation. 

Calculations were performed within multiple scattering theory using the muffin-tin potential approximation. The cluster radius was set to 6.4 \AA\;to include the full shell of $5^\mathrm{th}$ nearest neighbor Ru sites. The potential and Fermi level were determined self-consistently. Spin-polarized calculations were performed to determine the resonant magnetic scattering contributions while enforcing a net spin moment of 2$\mu_B$/Ru unless otherwise stated. All calculations exclude valence band spin-orbit coupling. 

For comparison to the experimental data in Fig. \ref{fig:fig3}(b), a self-absorption correction was applied to the calculated resonance spectrum as implemented in the FDMNES program \cite{Bunau2009}. The polarization averaged absorption coefficient is calculated $\bar{\mu}$ considering all atoms. The corrected intensity vs. $E_i$ ($I_c$) is then calculated from the uncorrected intensity ($I_{nc}$) as $I_c = \frac{\mu_0}{\bar{\mu}} I_{nc}$ where $\mu_0$ is the absorption in the pre-edge region \cite{Joly2012}. No corrections were applied to the experimental data.
 \\

\end{document}